\documentclass[sigconf,balance=false]{acmart}
\usepackage{algorithm}
\usepackage{algpseudocode}
\usepackage{multirow}

\setcopyright{none}
\acmConference[WWW 2027 submission]{The ACM Web Conference 2027}{2027}{Dublin, Ireland}
\acmYear{2027}
\acmDOI{}
\acmISBN{}

\hypersetup{pdfauthor={Manzong Huang, Chenyang Bu, Yi He, Jiapu Wang, Xindong Wu}}

\begin{document}

\title[QueryFlow: Multi-Hop Graph Retrieval with Adaptive Query Evolution]{QueryFlow: Multi-Hop Graph Retrieval with Adaptive Query Evolution}

\author{Manzong Huang}
\affiliation{
  \department[0]{Key Laboratory of Knowledge Engineering with Big Data (the Ministry of Education of China)}
  \department[1]{School of Computer Science and Information Engineering}
  \institution{Hefei University of Technology}
  \city{Hefei}
  \country{China}
}
\email{manzonghuang@mail.hfut.edu.cn}

\author{Chenyang Bu}
\authornote{Corresponding authors: Chenyang Bu and Xindong Wu.}
\affiliation{
  \department[0]{Key Laboratory of Knowledge Engineering with Big Data (the Ministry of Education of China)}
  \department[1]{School of Computer Science and Information Engineering}
  \institution{Hefei University of Technology}
  \city{Hefei}
  \country{China}
}
\email{chenyangbu@hfut.edu.cn}

\author{Yi He}
\affiliation{
  \department{Department of Data Science}
  \institution{College of William and Mary}
  \city{Williamsburg}
  \state{Virginia}
  \country{USA}
}
\email{yihe@wm.edu}

\author{Jiapu Wang}
\affiliation{
  \institution{Nanjing University of Science and Technology}
  \city{Nanjing}
  \country{China}
}
\email{jiapu.wang@njust.edu.cn}

\author{Xindong Wu}
\authornotemark[1]
\affiliation{
  \department[0]{Key Laboratory of Knowledge Engineering with Big Data (the Ministry of Education of China)}
  \department[1]{School of Computer Science and Information Engineering}
  \institution{Hefei University of Technology}
  \city{Hefei}
  \country{China}
}
\email{xwu@hfut.edu.cn}

\renewcommand{\shortauthors}{Huang et al.}

\begin{abstract}
Multi-hop retrieval-augmented generation (RAG) supports Web information seeking and fact verification by recovering complementary evidence across related pages within a limited retrieval budget. Scoring evidence independently against the original question can miss supporting units whose relevance emerges through intermediate facts. Iterative retrieval incorporates these facts into subsequent queries, while graph retrieval explicitly models relations among evidence units. Yet fixed-query graph retrievers update entity representations without adapting the query condition to accumulated relational context. We view multi-hop evidence recovery as the coupled evolution of evidence representations and query conditions, with accumulated context informing both what the retriever represents and how it searches for complementary evidence. QueryFlow realizes this perspective by recurrently integrating query-conditioned summaries of graph context into a shared query state, which controls boundary inputs and relation features for subsequent propagation. Across four representative multi-hop question answering (QA) datasets, QueryFlow achieves macro Recall@10 (R@10), Full Recall@10 (FR@10), and LLM-Acc gains of 4.98, 8.40, and 8.0 percentage points over representative baselines, respectively.
\end{abstract}


\maketitle

\section{Introduction}

\begingroup
\emergencystretch=2em
Retrieval-augmented generation (RAG) grounds language-model answers in external evidence~\cite{lewis2020rag}. In Web content mining, it supports complex information seeking and fact verification across related pages~\cite{talmor-berant-2018-web,jiang-etal-2020-hover}. Multi-hop questions make this evidence requirement more demanding because their answers depend on linked facts distributed across pages about related entities~\cite{ho-etal-2020-constructing,trivedi-etal-2022-musique,yang-etal-2018-hotpotqa}. Answering these questions requires retrieving evidence units, such as documents, passages, or chunks, that jointly support the necessary reasoning steps. A collection of individually relevant units can still omit a fact needed to complete the reasoning chain. Multi-hop retrieval is fundamentally an evidence-set recovery problem, with the goal of recovering the complete supporting set within a given retrieval budget. By improving multi-hop question answering over interconnected Web content, our work aligns with the \textbf{Web Mining, Multimedia and Multilingual Content Analysis track.}

\begin{figure}[t]
    \centering
    \includegraphics[width=1.0\columnwidth]{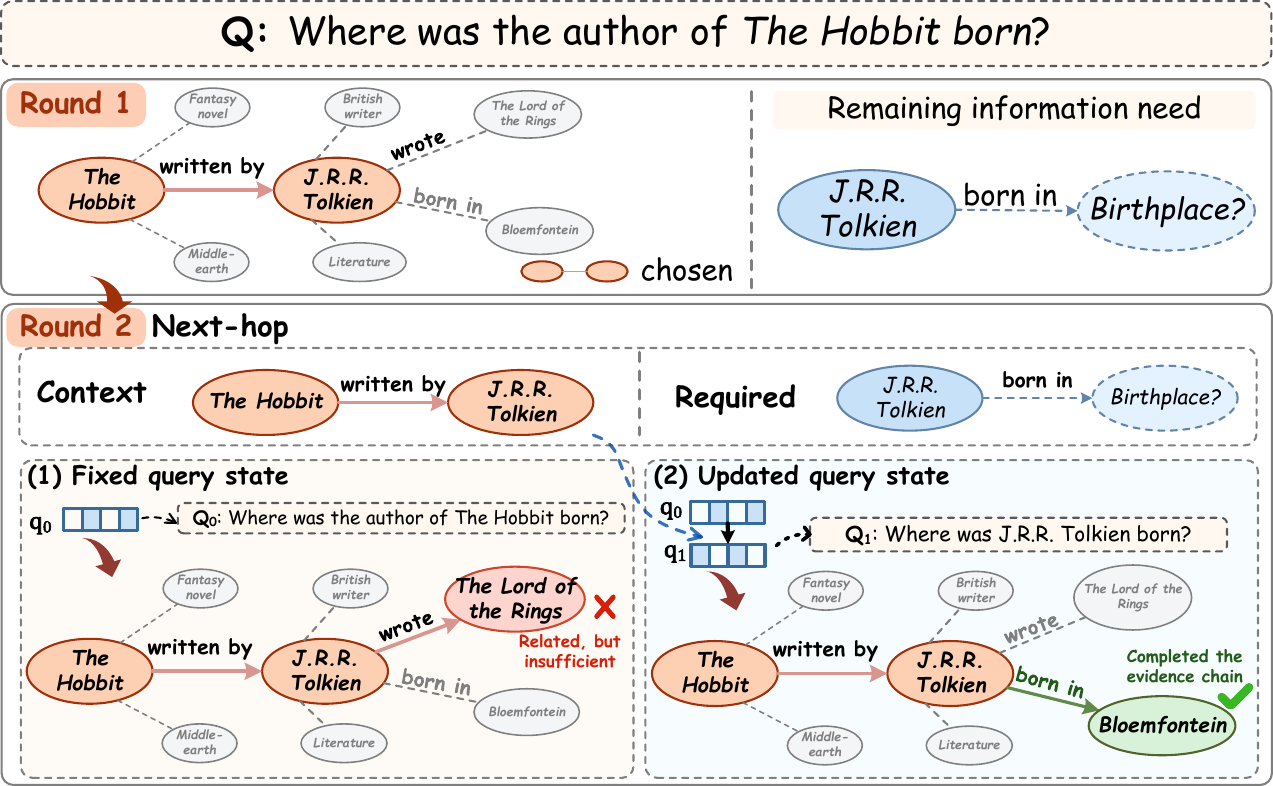}
    \caption{A fixed query state continues to favor evidence directly related to the original question. Updating the state with accumulated context guides retrieval toward complementary evidence, helping complete the support set. $Q_1$ illustrates the updated state's meaning.}
    \Description{A two-round retrieval example compares fixed and updated query states. After identifying J. R. R. Tolkien as the author of The Hobbit, the fixed query state continues to favor related book evidence. Updating the query state with this context guides retrieval toward Tolkien's birthplace, completing the supporting evidence in the example. Q1 illustrates the meaning of the updated latent query state.}
    \label{fig:motivation}
\end{figure}

Recovering a complete supporting set requires more than matching individual evidence units to the original question. Single-round dense retrievers learn semantic matches between questions and candidate evidence~\cite{karpukhin2020dpr}, but independent scoring can miss complementary units whose relevance becomes apparent only after intermediate facts are found. Iterative retrieval addresses this limitation by conditioning subsequent searches on retrieved passages or intermediate reasoning~\cite{xiong2021mdr,trivedi-etal-2023-interleaving}. In these approaches, evidence dependencies are conveyed through retrieval history, leaving entity relations across documents implicit. Graph-based RAG makes these connections explicit through entities and relations~\cite{graphrag,zhu-etal-2025-knowledge}, while neural graph retrievers learn to integrate connected evidence through query-conditioned message passing~\cite{li2025gnnret,luo2025gfmrag}. A central question in learned graph retrieval is how accumulated relational context should guide subsequent evidence acquisition. Fixed-query designs update entity representations while retaining the original query condition~\cite{luo2025gfmrag}. As illustrated in Figure~\ref{fig:motivation}, identifying the author of \textit{The Hobbit} shifts the remaining information need toward that person's birthplace, while a fixed query can continue to favor book-related evidence. These designs accumulate relational context in entity representations without an explicit feedback path for that context to revise the query condition governing subsequent propagation.

The query state should evolve with the evidence representations it guides. Intermediate entities and relations can clarify what remains to be retrieved, so accumulated graph context should inform both the representation of evidence and the condition used to explore further evidence. Coupling these updates brings relational propagation and query adaptation into the same retrieval process, allowing newly established context to guide the recovery of missing support.

We introduce \textbf{QueryFlow}, a graph retrieval framework that couples recurrent query evolution with evidence propagation to support multi-hop question answering. It maintains a shared query state alongside entity representations that accumulate relational context through message passing. Between successive propagation layers, query-conditioned attention summarizes all entity representations, making context distributed across the graph available to the query update. A gated recurrent transition integrates this summary with the preceding query state to revise the retrieval condition. Boundary feedback injects the updated state at question-linked entities, while relation feedback modulates the features used to transform neighboring messages. These interfaces change both the context injected into the graph and how evidence propagates across its relations, closing the feedback path from accumulated evidence to subsequent retrieval. Finally, entity scores are aggregated into a document ranking that supplies evidence for answer generation.

On four representative multi-hop QA datasets, QueryFlow achieves macro R@10, FR@10, and LLM-Acc gains of 4.98, 8.40, and 8.0 percentage points over representative baselines, respectively. Improvements in evidence coverage and complete recovery are accompanied by higher downstream QA reasoning accuracy, supporting the value of complete evidence retrieval for multi-hop question answering.

Our contributions are threefold.
\begin{itemize}
    \item We frame multi-hop graph retrieval as coupled evolution of evidence representations and query conditions, treating accumulated relational context as feedback for subsequent evidence acquisition.
    \item We propose \textbf{QueryFlow}, which couples recurrent query updates with graph propagation through boundary and relation feedback.
    \item Experiments on four multi-hop QA benchmarks show gains over representative baselines in evidence recovery and answer accuracy.
\end{itemize}
\par\endgroup

\section{Related Work}
\label{sec:related-work}
\begingroup
\tolerance=200
\emergencystretch=2em

\textbf{Relevance-based retrieval.}
RAG integrates external evidence into language-model training or generation~\cite{lewis2020rag,guu2020realm,borgeaud2022retro,izacard2023atlas}. Dense representations and token interactions improve matching between questions and evidence~\cite{karpukhin2020dpr,izacard2022contriever,khattab2020colbert,colbertv2}. Precomputed passage representations support efficient search, while token-level interactions retain fine-grained matching signals within each candidate passage. Better query representations, evidence fusion, and hierarchical indexing further improve evidence access and use~\cite{hyde,izacard2021fid,shi-etal-2024-replug,raptor}. In single-round matching, retrieved evidence does not update the scoring condition, so complementary support with weak direct relevance can be missed. QueryFlow complements relevance scoring with feedback from accumulated relational context.

\textbf{Iterative and adaptive retrieval.}
Iterative retrievers use retrieved passages, condensed facts, or rewritten queries to guide later searches~\cite{qi-etal-2019-answering,zhang-etal-2024-end,ma-etal-2023-query}. MDR re-encodes the question with earlier passages~\cite{xiong2021mdr}, while Baleen condenses retrieved passages into facts for later hops~\cite{khattab2021baleen}. Intermediate generation and reasoning also supply feedback~\cite{shao-etal-2023-enhancing,trivedi-etal-2023-interleaving}, while adaptive approaches control when and how to retrieve based on generation state or question difficulty~\cite{jiang-etal-2023-active,su-etal-2024-dragin,asai2024selfrag,jeong-etal-2024-adaptive}. In retrieval loops driven by passages or reasoning, feedback passes through text history, leaving entity relations across documents implicit. QueryFlow brings this adaptation into graph propagation, using intermediate entity context to update the shared retrieval condition.

\textbf{Graph-based retrieval.}
Knowledge graph research also studies graph completion, which predicts missing entities or relations from graph structure. Completion methods learn multi-concept and fine-grained entity--relation representations~\cite{wang2023mcrl,wang2024mli}, or adapt how entity, relation, and temporal information is combined to predict missing facts~\cite{wang2023tdn,wang2024ime,wang2024qdn}. Broader graph representation learning combines local and global signals for node classification, while attributed graph clustering uses collaborative graph augmentation and contrastive learning~\cite{wang2026comgrl,zhao2025ragc}. These tasks learn graph representations or predict missing facts, whereas graph retrieval ranks source documents to recover evidence supporting a question.
Graph indices and diffusion associate evidence through explicit links~\cite{graphrag,lightrag,zhu-etal-2025-knowledge,hipporag,hipporag2}, while subgraph selection and path exploration construct structured context~\cite{g-retriever,hu2025grag,li2025subgraphrag,sun2024tog,luo2024rog,luo2026graphr1}. Learned relational reasoning supports query-conditioned propagation~\cite{zhu2021nbfnet,zhu2023astarnet,galkin2024ultra,zhuo2026krlm,wang2024llmda}, and neural retrievers learn to aggregate or select connected evidence~\cite{sun2019pullnet,mavromatis2025gnnrag,li2025gnnret}. Learned graph retrievers differ in how relational context conditions retrieval. GFM-RAG propagates entity relevance under a fixed query and maps it to documents~\cite{luo2025gfmrag}, whereas D-RAG updates instructions from graph context for knowledge graph question answering~\cite{gao-etal-2025-rag,zhuo2025effective,ijcai2025p413}. QueryFlow studies this interaction for complete document evidence recovery, using a shared evolving query state to control boundary inputs and relation features.

\par\endgroup

\section{Method}
\label{sec:method}

\begin{figure*}[t]
    \centering
    \includegraphics[width=0.99\textwidth]{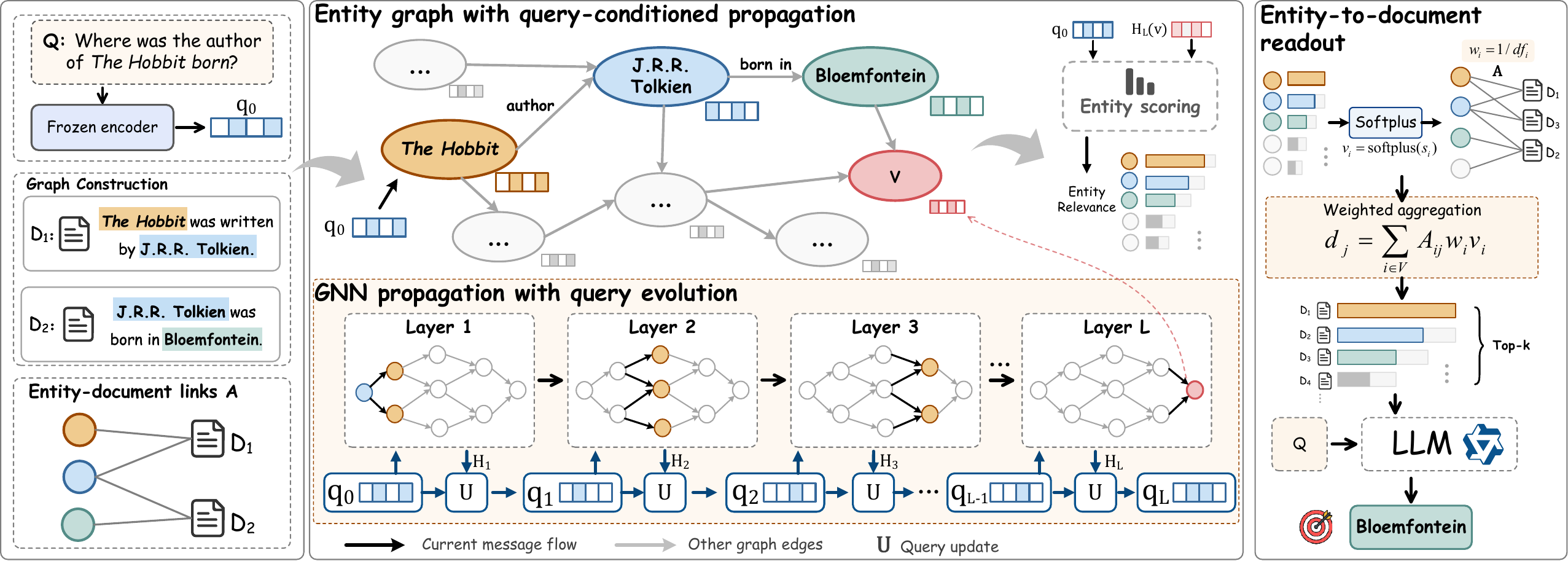}
    \caption{Overview of QueryFlow for multi-hop evidence retrieval from interconnected document collections. A question is grounded in an entity graph whose nodes are linked to source documents. Graph propagation and recurrent query updates alternate across layers; the updated query state conditions the boundary inputs and relation features of the next layer. Final entity representations are scored together with the original query, and the scores of all entities are transformed by softplus and aggregated through associations between entities and documents using reciprocal document frequency weights. The documents with the highest scores provide evidence for answer generation. The terminal $q_L$ update is used only for the readout ablation.}
    \Description{The question initializes a query vector and linked entity states. Intermediate graph layers send entity representations to recurrent query updates that control the next layer. All entity representations are scored with the original query. Softplus scores, reciprocal document frequencies, and links between entities and documents yield document scores; the documents with the highest scores support answer generation. The terminal update after the final graph layer produces qL only for the readout ablation.}
    \label{fig:method}
\end{figure*}

\textbf{Overview.}
QueryFlow organizes multi-hop evidence retrieval as a feedback process between entity representations and a shared query state (Figure~\ref{fig:method}). Entity propagation aggregates relational context, while recurrent query evolution converts that context into the condition for subsequent propagation. The evolving query state controls both boundary inputs and relation features, allowing accumulated context to influence how further evidence is explored. Final entity scores are aggregated into a document ranking for answer generation.

\subsection{Evidence Graph and Query Grounding}
\label{sec:formulation}

\textbf{Retrieval task and graph interface.}
Given a question $x$ and a document corpus $D$, the objective is to recover the set of supporting documents $D^*(x)\subseteq D$ within a retrieval budget $k$. The first $k$ documents in the output ranking form $\widehat D_k(x)$. We represent the corpus by an entity graph $G=(V,E,\mathcal R)$~\cite{luo2025gfmrag}, where $V$ contains entities and $E$ contains edges labeled by relation types in $\mathcal R$. An incidence matrix $A\in\{0,1\}^{|V|\times|D|}$ records associations between entities and documents, with $A_{ij}=1$ when entity $i$ occurs in document $j$. Message passing accumulates context at entity nodes; the incidence matrix subsequently transfers entity relevance to documents.

\textbf{Query and entity initialization.}
A frozen text encoder produces question and relation embeddings $e_x$ and $e_r$. Learned linear projections $\phi_q$ and $\phi_r$ map them to a common $d$-dimensional space. Let $m_x\in\{0,1\}^{|V|}$ identify the entities linked to $x$. The initial query, relation features, and entity states are
\begin{equation}
    \begin{aligned}
        q_0&=\phi_q(e_x),\qquad R^0(r)=\phi_r(e_r),\\
        H_0(i)&=m_x(i)q_0.
    \end{aligned}
    \label{eq:initialization}
\end{equation}
Thus entities linked to the question start from $q_0$, while other entities start from zero. At layer $l$, the shared query state $q_{l-1}\in\mathbb R^d$ controls the update to entity representations $H_l\in\mathbb R^{|V|\times d}$. The entity representations store context specific to each entity, while the query supplies a shared condition for propagation. Graph topology, document associations, and the grounding mask remain fixed as these representations evolve.

\subsection{Query-Conditioned Evidence Propagation}
\label{sec:propagation}
\label{sec:control}

Each layer propagates entity context using boundary inputs and relational messages~\cite{zhu2021nbfnet}. The shared query state $q_{l-1}$ controls both the representations injected at question-linked entities and the relation features that transform neighboring messages.

\textbf{Boundary feedback.}
The boundary input at the original grounding locations is
\begin{equation}
    B_l(i)=m_x(i)q_{l-1}.
    \label{eq:boundary}
\end{equation}
Its support is fixed by $m_x$, while its content changes with the query state. At later layers, this input reintroduces the context encoded by earlier query updates, maintaining a connection between question grounding and evolving relational information.

\textbf{Relation feedback.}
The same state generates a gate for each feature of the base relation embeddings,
\begin{equation}
    \begin{aligned}
        \gamma_{l-1}&=\sigma(W_rq_{l-1}+b_r),\\
        \bar R_l(r)&=\psi_l\!\left(R^0(r)\odot\gamma_{l-1}\right).
    \end{aligned}
    \label{eq:relation-control}
\end{equation}
Here $\sigma$ is the sigmoid function, $\odot$ denotes elementwise multiplication, and $\psi_l$ is a layer-specific two-layer MLP with an intervening ReLU. The gate is shared across relation types, each retaining its own base embedding $R^0(r)$. It modulates dimensions of the relation features according to the current query condition before the layer-specific transformation.

\textbf{Relational message passing.}
Let $\mathcal N(i)$ contain pairs of source entities and relations $(j,r)$ with an edge from $j$ to $i$. DistMult messages~\cite{yang2015distmult} and sum aggregation combine neighboring states with the boundary to give
\begin{equation}
    u_{l,i}=B_l(i)+\sum_{(j,r)\in\mathcal N(i)}
        H_{l-1}(j)\odot\bar R_l(r).
    \label{eq:messages}
\end{equation}
The next entity representation is obtained by transforming the concatenated previous state and aggregate, followed by a residual connection:
\begin{equation}
    \begin{aligned}
        t_{l,i}&=W_l[H_{l-1}(i);u_{l,i}]+c_l,\\
        H_l(i)&=H_{l-1}(i)+
            \operatorname{ReLU}(\operatorname{LN}_l(t_{l,i})).
    \end{aligned}
    \label{eq:node-update}
\end{equation}
Here $[\,;\,]$ denotes concatenation, and $\operatorname{LN}_l$ is layer-specific normalization. Boundary feedback changes the representation injected at the starting entities, while relation feedback changes the features used to carry messages along edges. For $l<L$, $H_l$ supplies the next layer's node input and the context for query evolution.

\subsection{Graph-Conditioned Query Evolution}
\label{sec:transition}

Query evolution summarizes $H_l$ to update the shared state $q_l$. For $l<L$, this state conditions the next propagation layer. The readout ablation additionally applies the same update at $l=L$ to $(H_L,q_{L-1})$, obtaining the terminal state $q_L$ without further graph propagation.

\textbf{Query-conditioned graph summary.}
We use the preceding query state to attend to the new entity representations with weights
\begin{equation}
    a_{l,i}=\frac{\exp\!\left(H_l(i)^\top q_{l-1}/\sqrt d\right)}
        {\sum_{j\in V}\exp\!\left(H_l(j)^\top q_{l-1}/\sqrt d\right)}.
    \label{eq:attention}
\end{equation}
The normalized graph summary is
\begin{equation}
    z_l=\operatorname{LN}_{\mathrm{sum}}
        \left(\sum_{i\in V}a_{l,i}H_l(i)\right).
    \label{eq:summary}
\end{equation}
The softmax ranges over all entities in $V$. Consequently, $z_l$ depends jointly on the current query and the graph representations, emphasizing context compatible with the retrieval condition. The summary provides an input of fixed width to the controller regardless of graph size.

\textbf{Gated recurrent transition.}
A gated recurrent unit (GRU)~\cite{cho-etal-2014-learning} integrates $z_l$ with the preceding query state. An additional gate regulates the displacement in each feature dimension toward the candidate:
\begin{equation}
    \begin{aligned}
        \widetilde q_l&=\operatorname{GRUCell}(z_l,q_{l-1}),\\
        g_l&=\sigma\!\left(W_g[q_{l-1};z_l]+b_g\right),\\
        \Delta_l&=g_l\odot(\widetilde q_l-q_{l-1}).
    \end{aligned}
    \label{eq:state-gate}
\end{equation}
The candidate $\widetilde q_l$ combines previous query information with new graph context, while $g_l$ controls how much of their difference enters the update. A scalar $\alpha\in[0,1]$ scales this increment before query normalization:
\begin{equation}
    q_l=
    \begin{cases}
        q_{l-1}, & \alpha=0,\\
        \operatorname{LN}_{q}(q_{l-1}+\alpha\Delta_l),
            & 0<\alpha\leq1.
    \end{cases}
    \label{eq:transition}
\end{equation}
We use $\alpha=1$ for recurrent propagation. Setting $\alpha=0$ retains $q_l=q_0$ by bypassing the transition and its normalization, providing a fixed-query intervention within the same model. The recurrent cell, transition gate, and relation gate share parameters across layers; $\operatorname{LN}_{\mathrm{sum}}$ and $\operatorname{LN}_q$ have separate parameters, each shared across layers. For $l<L$, the updated state controls the next boundary and relation features through Eqs.~\eqref{eq:boundary} and~\eqref{eq:relation-control}, completing the feedback from graph context to subsequent propagation.

\subsection{Document Readout and Learning}
\label{sec:readout}

\textbf{Entity relevance scoring.}
After $L$ propagation layers, a shared two-layer MLP with an intervening ReLU scores each entity using its final representation and the initial query:
\begin{equation}
    s_i=\operatorname{MLP}\!\left([H_L(i);q_0]\right).
    \label{eq:entity-score}
\end{equation}
The default model uses $q_0$ in both training and retrieval to anchor scores to the original question. The readout ablation replaces $q_0$ with $q_L$ in Eq.~\eqref{eq:entity-score}, keeping $H_L$ unchanged. The entity-level objective trains the query controller through its influence on subsequent graph propagation.

\textbf{Entity-to-document readout.}
We convert each entity logit $s_i$ into a positive contribution weight using
\begin{equation}
    v_i=\operatorname{softplus}(s_i)=\log(1+\exp(s_i)).
    \label{eq:softplus}
\end{equation}
Let $f_i=\sum_{j=1}^{|D|}A_{ij}$ denote the document frequency of entity $i$. We assign reciprocal frequency weights~\cite{luo2025gfmrag},
\begin{equation}
    \begin{aligned}
        w_i^{\mathrm{doc}}&=
            \begin{cases}
                f_i^{-1}, & f_i>0,\\
                0, & f_i=0,
            \end{cases}\\
        d_j&=\sum_{i\in V}A_{ij}\,w_i^{\mathrm{doc}}v_i.
    \end{aligned}
    \label{eq:document-score}
\end{equation}
The readout aggregates contributions from all entities associated with each document. For entities that occur in many documents, reciprocal frequency weighting reduces their contribution to each document. We rank documents by $d_j$ and return the top $k$ as $\widehat D_k(x)$.

\textbf{Entity-level learning.}
Let $y_i\in\{0,1\}$ be an entity supporting label and $P=\{i:y_i=1\}$ the positive set. With $p_i=\sigma(s_i)$, the listwise loss that assigns normalized score mass to each supporting entity is
\begin{equation}
    \begin{aligned}
        \pi_i&=\frac{p_i}{\sum_{j\in V}p_j+\varepsilon},\\
        \mathcal L_{\mathrm{ListCE}}
           &=-\frac{1}{|P|}\sum_{i\in P}\log(\pi_i+\varepsilon).
    \end{aligned}
    \label{eq:listce}
\end{equation}
Here $\varepsilon$ is a numerical stabilizer. We combine the listwise objective with a weighted binary cross-entropy term that classifies individual entities and emphasizes negatives with high scores to obtain
\begin{equation}
    \mathcal L_x=0.3\,\mathcal L_{\mathrm{BCE}}
                  +0.7\,\mathcal L_{\mathrm{ListCE}}.
    \label{eq:loss}
\end{equation}
Gradients from entity scores jointly train graph propagation and the recurrent controller. Aggregation of entity scores into document scores and document ranking are applied at inference. Appendix~\ref{app:implementation} details the training objective and retrieval algorithm.

\begin{table*}[t]
    \centering
    \caption{Answer generation across four datasets. Scores range from 0 to 1. Baselines are grouped by retrieval mechanism and ordered within each group by average LLM-Acc. Averages are unweighted. Best in \textbf{bold}, second \underline{underlined}.}
    \label{tab:answer-results}
    \small
    \renewcommand{\arraystretch}{1.05}
    \setlength{\tabcolsep}{5pt}
    \begin{tabular}{@{}ll*{10}{c}@{}}
        \toprule
        \multirow{2}{*}{Category} & \multirow{2}{*}{Method}
        & \multicolumn{2}{c}{2WikiMultiHopQA}
        & \multicolumn{2}{c}{HotpotQA}
        & \multicolumn{2}{c}{MuSiQue}
        & \multicolumn{2}{c}{ConcurrentQA}
        & \multicolumn{2}{c}{Average} \\
        \cmidrule(lr){3-4}\cmidrule(lr){5-6}\cmidrule(lr){7-8}
        \cmidrule(lr){9-10}\cmidrule(lr){11-12}
        & & LLM-Acc & Contain & LLM-Acc & Contain & LLM-Acc & Contain
        & LLM-Acc & Contain & LLM-Acc & Contain \\
        \midrule
        \multirow{3}{*}{\shortstack[l]{Query-based\\retrieval}}
        & Dense RAG & 0.235 & 0.228 & 0.592 & 0.517 & 0.241 & 0.185 & 0.331 & 0.381 & 0.350 & 0.328 \\
        & HyDE & 0.273 & 0.266 & 0.654 & 0.568 & 0.288 & 0.231 & 0.374 & 0.415 & 0.397 & 0.370 \\
        & ColBERTv2 & 0.313 & 0.306 & 0.624 & 0.538 & 0.237 & 0.187 & 0.464 & 0.486 & 0.410 & 0.379 \\
        \midrule
        \multirow{4}{*}{\shortstack[l]{Iterative and\\structure-aware\\retrieval}}
        & FastToG & 0.326 & 0.331 & 0.385 & 0.327 & 0.077 & 0.068 & 0.233 & 0.269 & 0.255 & 0.249 \\
        & RAPTOR & 0.264 & 0.269 & 0.393 & 0.346 & 0.102 & 0.076 & 0.308 & 0.364 & 0.267 & 0.264 \\
        & ToG & 0.418 & 0.419 & 0.317 & 0.282 & 0.111 & 0.083 & 0.221 & 0.196 & 0.267 & 0.245 \\
        & PathRAG & 0.443 & 0.426 & 0.465 & 0.390 & 0.163 & 0.125 & 0.203 & 0.185 & 0.319 & 0.282 \\
        \midrule
        \multirow{7}{*}{\shortstack[l]{Graph-based\\retrieval}}
        & G-Retriever & 0.083 & 0.099 & 0.322 & 0.284 & 0.106 & 0.077 & 0.216 & 0.299 & 0.182 & 0.190 \\
        & LightRAG & 0.188 & 0.188 & 0.535 & 0.453 & 0.232 & 0.160 & 0.293 & 0.339 & 0.312 & 0.285 \\
        & GraphRAG & 0.358 & 0.357 & 0.484 & 0.437 & 0.140 & 0.111 & 0.357 & 0.381 & 0.335 & 0.322 \\
        & DALK & 0.384 & 0.363 & 0.463 & 0.356 & 0.213 & 0.140 & 0.338 & 0.305 & 0.350 & 0.291 \\
        & HippoRAG 2 & 0.447 & 0.580 & 0.701 & 0.644 & 0.291 & 0.283 & 0.485 & 0.549 & 0.481 & 0.514 \\
        & GFM-RAG & 0.668 & \underline{0.744} & 0.718 & \underline{0.657} & 0.320 & 0.318 & 0.613 & \underline{0.625} & 0.580 & \underline{0.586} \\
        & LinearRAG & \underline{0.679} & 0.694 & \underline{0.752} & 0.647 & \underline{0.369} & \underline{0.332} & \underline{0.662} & 0.617 & \underline{0.616} & 0.573 \\
        \midrule
        Proposed & QueryFlow & \textbf{0.829} & \textbf{0.819} & \textbf{0.786} & \textbf{0.695} & \textbf{0.439} & \textbf{0.375} & \textbf{0.732} & \textbf{0.678} & \textbf{0.696} & \textbf{0.642} \\
        \bottomrule
    \end{tabular}
\end{table*}

\section{Experiments}

Our experiments connect answer quality to the recovery of supporting evidence. We first compare generation and retrieval performance, then examine query evolution and its feedback interfaces through controlled interventions. Efficiency, retrieval budgets, and paired recovery analyses characterize the costs and conditions under which these gains occur.

\subsection{Experimental Setup}
\label{sec:exp-settings}

\textbf{Datasets.}
We evaluate on four multi-hop QA benchmarks, namely 2WikiMultiHopQA~\cite{ho-etal-2020-constructing}, HotpotQA~\cite{yang-etal-2018-hotpotqa}, MuSiQue~\cite{trivedi-etal-2022-musique}, and ConcurrentQA~\cite{concurrentqa}. Their annotated supporting documents define the target evidence sets. Dataset and evaluation details are provided in Appendix~\ref{app:experimental-setup}. Model and training settings appear in Appendix~\ref{app:implementation-settings}.

\textbf{Evaluation metrics.}
For answer generation, we report LLM-Acc, the fraction of answers accepted by an LLM judge, and Contain, the fraction of generated answers containing the normalized gold answer string. For retrieval budget $k$, document recall and full recall are
\begin{equation}
    \begin{aligned}
        \operatorname{R@}k(x)
        &=\frac{|\widehat D_k(x)\cap D^*(x)|}{|D^*(x)|},\\
        \operatorname{FR@}k(x)
        &=\mathbb I[D^*(x)\subseteq\widehat D_k(x)],
    \end{aligned}
    \label{eq:metrics}
\end{equation}
where $\mathbb I$ is the indicator function. Both metrics are averaged over questions with nonempty support sets; FR@$k$ measures complete recovery of the supporting set. We report results for each dataset and an unweighted macro average at $k=10$ and $k=30$ (Tables~\ref{tab:macro-retrieval-results} and~\ref{tab:retrieval-results-appendix}).

\textbf{Retrieval comparisons.}
We compare R@$k$ and FR@$k$ only for methods whose final retrieval output is a ranked list of documents. Methods returning only graph structures (e.g., ToG~\cite{sun2024tog}) or text chunks without a document ranking are evaluated through downstream answer quality instead.

\label{sec:exp-baselines}
\begingroup
\tolerance=200
\emergencystretch=2em
\textbf{Baselines.}
Our retrieval baselines cover hierarchical retrieval with RAPTOR~\cite{raptor}, query expansion with HyDE~\cite{hyde}, and late-interaction retrieval with ColBERTv2~\cite{colbertv2}. The six graph retrieval baselines are G-Retriever~\cite{g-retriever}, LightRAG~\cite{lightrag}, GraphRAG~\cite{graphrag}, DALK~\cite{dalk}, HippoRAG~2~\cite{hipporag2}, and GFM-RAG~\cite{luo2025gfmrag}. Appendix~\ref{app:additional-results} provides retrieval results for each dataset.
\par\endgroup

\textbf{Answer generation.}
We use Qwen3.5-9B for answer generation across all methods. Following LinearRAG~\cite{linearrag}, we report LLM-Acc and Contain using the definitions above. Table~\ref{tab:answer-results} compares the retrieval methods used to supply evidence, including RAPTOR, HyDE, ColBERTv2, G-Retriever, LightRAG, GraphRAG, DALK, GFM-RAG, LinearRAG, Dense RAG, FastToG~\cite{fasttog}, Think-on-Graph (ToG)~\cite{sun2024tog}, PathRAG~\cite{pathrag}, and HippoRAG~2~\cite{hipporag2}. The table groups these methods by retrieval mechanism.

\subsection{Main Results}
\label{sec:main-experiments}

Table~\ref{tab:answer-results} compares downstream answer quality across query-based, iterative and structure-aware, and graph-based retrieval methods on four benchmarks. These results address the following research question.
\par\nopagebreak[4]

\noindent\textbf{RQ1.} \emph{How does QueryFlow compare with existing retrieval and reasoning methods on downstream answer generation?}

\textbf{Answer quality improves over representative baselines.}
QueryFlow reaches macro LLM-Acc of 0.696 and Contain of 0.642. Relative to LinearRAG, these scores increase by 8.0 and 6.9 percentage points, respectively. The gains over GFM-RAG are 11.6 and 5.6 points. Both comparisons show improvements in judged correctness and inclusion of the gold answer.

\textbf{The answer gains extend across all four benchmarks.}
Compared with LinearRAG, LLM-Acc increases by 15.0 points on 2WikiMultiHopQA, 3.4 on HotpotQA, and 7.0 on both MuSiQue and ConcurrentQA. On MuSiQue, LLM-Acc rises from 0.369 to 0.439 while Contain improves from 0.332 to 0.375. QueryFlow improves both metrics on every dataset relative to LinearRAG and GFM-RAG, showing that the macro gains extend across all four collections.

\textbf{Semantic correctness and answer inclusion improve together.}
Contain gains range from 4.3 to 12.5 points over LinearRAG and from 3.8 to 7.5 points over GFM-RAG. For multi-hop questions, intermediate facts identify the entity or relation to which the final answer must refer. QueryFlow feeds this context into subsequent propagation to guide retrieval of complementary support. The gains in complete recovery examined in RQ2 are consistent with this explanation of the answer gains.

\begin{table}[!htbp]
    \centering
    \caption{Macro retrieval scores (\%), ordered by FR@30. We compare R@$k$ and FR@$k$ for methods with document-level outputs, obtained directly or by mapping retrieved graph or hierarchical units to source documents. Results for each dataset appear in Table~\ref{tab:retrieval-results-appendix}. Best in \textbf{bold}, second \underline{underlined}.}
    \label{tab:macro-retrieval-results}
    \small
    \setlength{\tabcolsep}{10pt}
    \begin{tabular}{lrrrr}
        \toprule
        Method & R@10 & FR@10 & R@30 & FR@30 \\
        \midrule
        G-Retriever & 36.22 & 16.72 & 38.58 & 17.75 \\
        DALK & 57.09 & 29.38 & 60.63 & 33.09 \\
        RAPTOR & 56.53 & 27.84 & 68.19 & 43.84 \\
        LightRAG & 61.66 & 37.28 & 69.67 & 47.70 \\
        GraphRAG & 65.72 & 41.58 & 74.65 & 54.27 \\
        HyDE & 66.52 & 43.17 & 75.16 & 54.73 \\
        ColBERTv2 & 71.20 & 46.91 & 78.63 & 58.37 \\
        HippoRAG 2 & 74.29 & 51.38 & 80.95 & 61.59 \\
        GFM-RAG & \underline{76.95} & \underline{61.98} & \underline{81.41} & \underline{68.15} \\
        \midrule
        QueryFlow & \textbf{81.93} & \textbf{70.38} & \textbf{87.12} & \textbf{77.78} \\
        \bottomrule
    \end{tabular}
\end{table}

\begin{figure}[t]
    \centering
    \includegraphics[width=\columnwidth,trim=0 1bp 0 3bp,clip]{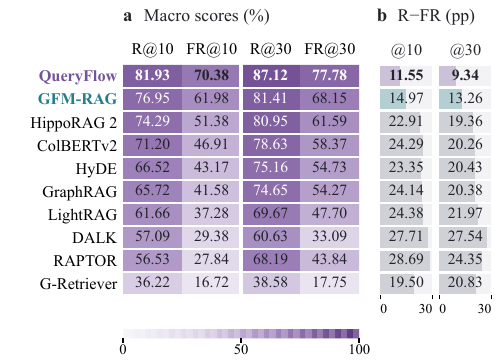}
    \caption{Macro retrieval comparison, ordered by R@10. (a) Scores from Table~\ref{tab:macro-retrieval-results}; darker cells indicate higher values. (b) Differences between the displayed R@$k$ and FR@$k$ at $k=10$ and $k=30$, with bars sharing a scale from 0 to 30 percentage points.}
    \Description{A score matrix shows R@10, FR@10, R@30, and FR@30 for all ten methods. Two aligned bar columns compare each method's gap between coverage and completeness at budgets ten and thirty on the same scale. QueryFlow scores 81.93, 70.38, 87.12, and 77.78 percent, with gaps of 11.55 and 9.34 percentage points, respectively.}
    \label{fig:retrieval-gap}
\end{figure}

To examine the evidence available for answer generation, we compare coverage of supporting documents and complete recovery of evidence sets at $k=10$ and $k=30$. Table~\ref{tab:macro-retrieval-results} and Figure~\ref{fig:retrieval-gap} summarize the results for the following question.
\par\nopagebreak[4]

\noindent\textbf{RQ2.} \emph{How does QueryFlow compare with existing retrieval methods on multi-hop evidence retrieval benchmarks?}

\textbf{Larger gains in complete recovery.}
QueryFlow improves macro FR@10 and FR@30 over GFM-RAG by 8.40 and 9.63 percentage points, reaching 70.38\% and 77.78\%, respectively (Table~\ref{tab:macro-retrieval-results}). Support coverage also improves, with R@10 of 81.93\% and R@30 of 87.12\%, gains of 4.98 and 5.70 points. The larger margins in full recall locate the main benefit in recovering every document required by a question. Retrieving the last missing supporting document turns partial coverage into complete support, so this objective favors complementary evidence. QueryFlow lets intermediate entity context revise the condition for later propagation, providing a mechanism for directing retrieval toward such evidence.

\textbf{Consistent gains across datasets.}
Across datasets at $k=10$, R@10 gains over GFM-RAG range from 4.57 to 5.28 points. The FR@10 gains are 12.60 points on 2WikiMultiHopQA, 8.11 on HotpotQA, 5.99 on MuSiQue, and 6.90 on ConcurrentQA (Table~\ref{tab:retrieval-results-appendix}). Each dataset exhibits a larger gain in full recall than in average support coverage. On MuSiQue, the margin in full recall grows from 5.99 points at $k=10$ to 8.88 at $k=30$, extending the improvement to the larger retrieval budget.

\textbf{Complete recovery distinguishes complementary support from average coverage.}
On MuSiQue, HippoRAG~2 has higher R@10 than QueryFlow (64.42\% versus 61.60\%), yet QueryFlow has higher FR@10 (37.82\% versus 31.11\%). The same ordering holds at $k=30$. Average coverage rewards each recovered document, whereas full recall requires all supporting documents to occur together. This contrast shows why coverage alone cannot characterize evidence-set recovery and why the two metrics provide complementary views of the retrieval objective.

\textbf{Closer agreement between coverage and completeness.}
Figure~\ref{fig:retrieval-gap} shows that QueryFlow's gap between support coverage and complete recovery decreases from 11.55 points at $k=10$ to 9.34 at $k=30$. GFM-RAG records gaps of 14.97 and 13.26 points, and HippoRAG~2 22.91 and 19.36 points, respectively. This difference captures the average recall contributed by questions whose support sets remain incomplete. Together with higher values of both metrics, these smaller gaps show that QueryFlow combines broader support coverage with more frequent complete recovery at both budgets.

\subsection{Query Evolution Analysis}
\label{sec:query-evolution-analysis}

\begin{figure}[htbp]
    \centering
    \includegraphics[width=\columnwidth,trim=0 9bp 0 5bp,clip]{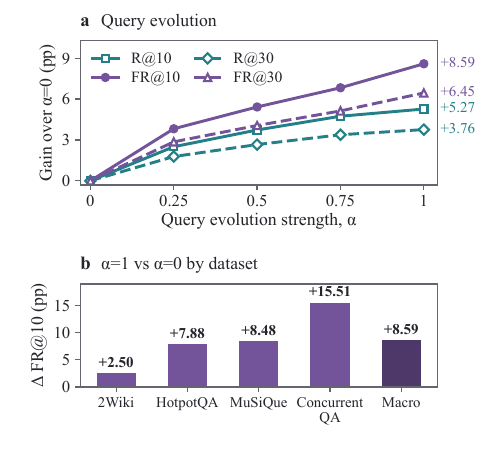}
    \caption{Effect of query evolution. (a) Macro recall and full-recall gains over $\alpha=0$ at $k=10$ and $k=30$. (b) FR@10 gains from $\alpha=0$ to $\alpha=1$ by dataset and their macro average with equal weights.}
    \Description{Panel a plots gains across alpha values zero, one quarter, one half, three quarters, and one. At alpha one, R@10, FR@10, R@30, and FR@30 improve by 5.27, 8.59, 3.76, and 6.45 percentage points. Panel b shows FR@10 gains of 2.50, 7.88, 8.48, and 15.51 points on 2WikiMultiHopQA, HotpotQA, MuSiQue, and ConcurrentQA, with a macro gain of 8.59 points.}
    \label{fig:query-evolution}
\end{figure}

We vary the query update strength $\alpha\in\{0,0.25,0.5,0.75,1\}$ on all four benchmarks while retaining the learned model parameters. Figure~\ref{fig:query-evolution} shows macro gains over propagation with a fixed query and the effects of enabling query evolution on each dataset.
\par\nopagebreak[4]

\noindent\textbf{RQ3.} \emph{How does adaptive query evolution affect retrieval performance?}

\textbf{Query evolution improves complete recovery across all four benchmarks.} At $\alpha=1$, macro R@10 and FR@10 reach 81.93\% and 70.38\%, gains of 5.27 and 8.59 percentage points over $\alpha=0$. Figure~\ref{fig:query-evolution}(b) shows positive FR@10 gains on every dataset, ranging from 2.50 points on 2WikiMultiHopQA to 15.51 on ConcurrentQA. With the anchored $q_0$ scorer and document readout fixed, query feedback affects document ranking through the propagated entity representations. The larger gain in full recall connects this feedback to recovering complementary support, beyond improving average coverage alone.

\textbf{Stronger updates consistently improve recovery across the tested settings.} From $\alpha=0.25$ to $\alpha=1$, macro R@10 increases from 79.16\% to 81.93\%, while FR@10 rises from 65.60\% to 70.38\% (Figure~\ref{fig:query-evolution}(a)). The same upward trend holds at $k=30$, where full updates gain 3.76 points in recall and 6.45 in full recall over $\alpha=0$. Increasing $\alpha$ gives the graph-derived update a larger role in the query state that controls subsequent boundary inputs and relation features. Continued improvement up to $\alpha=1$ supports using accumulated relational context throughout propagation.

\subsection{Ablation Study}
\label{sec:ablation-experiments}

Table~\ref{tab:ablation-results} examines the feedback interfaces and the choice between initial-query $q_0$ and terminal-query $q_L$ readout, retaining the fixed-query variant from RQ3 for reference. Appendix~\ref{app:readout-controls} examines entity aggregation alternatives.
\par\nopagebreak[4]

\noindent\textbf{RQ4.} \emph{How do boundary feedback, relation feedback, and the readout query contribute to retrieval performance?}

\begin{table}[tbp]
    \centering
    \caption{Macro ablation scores (\%). Best in \textbf{bold}.}
    \label{tab:ablation-results}
    \setlength{\tabcolsep}{4pt}
    \begin{tabular}{@{}lrrrr@{}}
        \toprule
        Variant & R@10 & FR@10 & R@30 & FR@30 \\
        \midrule
        \textbf{Full model} & \textbf{81.93} & \textbf{70.38} & \textbf{87.12} & \textbf{77.78} \\
        w/o query evolution ($\alpha=0$) & 76.66 & 61.78 & 83.36 & 71.33 \\
        w/o boundary feedback & 79.04 & 65.65 & 85.05 & 74.24 \\
        w/o relation feedback & 79.83 & 66.94 & 85.62 & 75.20 \\
        Readout with $q_L$ & 80.65 & 68.56 & 86.13 & 76.76 \\
        \bottomrule
    \end{tabular}
\end{table}

\textbf{Both interfaces contribute, with a larger effect from boundary feedback.}
Removing boundary feedback lowers FR@10 and FR@30 by 4.73 and 3.54 points, compared with 3.44 and 2.58 for relation feedback. The former refreshes query information at question-linked entities; the latter controls neighboring-message transformations. Their complementary roles support using both interfaces.

\textbf{Initial-query readout improves early ranking.}
The $q_0$ readout achieves R@10 and FR@10 of 81.93\% and 70.38\%, improving over $q_L$ by 1.28 and 1.82 percentage points (Table~\ref{tab:ablation-results}). At $k=30$, the corresponding scores are 87.12\% and 77.78\%, with gains of 0.99 and 1.02 points. The larger full-recall gain at $k=10$ supports anchoring final relevance scores to the original question while evolving queries guide context accumulation.

\subsection{Evidence Recovery Analysis}
\label{sec:evidence-recovery-analysis}

\begin{figure}[htbp]
    \centering
    \includegraphics[width=\columnwidth,trim=0 3bp 0 2bp,clip]{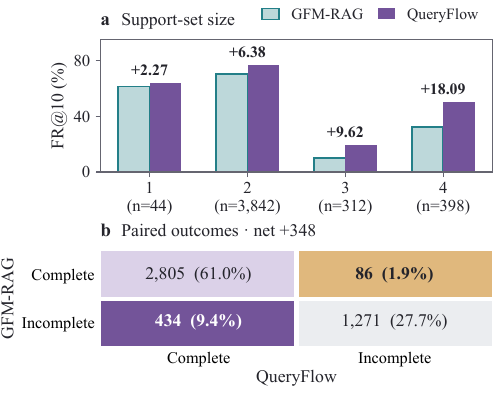}
    \caption{Evidence recovery at $k=10$. (a) FR@10 by support set size; labels show gains (pp), and $n$ is group size. (b) Paired outcomes. The 434 gains minus 86 losses yield 348 net recoveries.}
    \Description{Panel a compares GFM-RAG and QueryFlow full recall for support sets of size one through four. Panel b is an outcome matrix with two rows for GFM-RAG and two columns for QueryFlow. There are 2805 questions complete for both, 86 only for GFM-RAG, 434 only for QueryFlow, and 1271 incomplete for both, across 4596 paired questions.}
    \label{fig:evidence-recovery}
\end{figure}

Figure~\ref{fig:evidence-recovery} compares QueryFlow and GFM-RAG on 4,596 common evaluable questions using the shared support targets defined in Section~\ref{sec:exp-settings}. We report FR@10 by support set size and count paired changes between incomplete and complete recovery.
\par\nopagebreak[4]

\noindent\textbf{RQ5.} \emph{How do gains in complete recovery of evidence sets vary with support set size?}

\textbf{The largest absolute gain occurs for support sets with four documents.}
Figure~\ref{fig:evidence-recovery}(a) shows that QueryFlow raises FR@10 from 32.16\% to 50.25\% for support sets with four documents, a gain of 18.09 percentage points. The gains for sets with two and three documents are 6.38 and 9.62 points, respectively, extending the benefit to questions requiring several supporting documents. Larger support sets require more complementary documents to appear together, and a single missing document prevents full recovery. The larger margin in this group is consistent with feedback from graph context helping connect evidence distributed across multiple entities and documents. Support set size describes evidence requirements across the pooled benchmarks and also reflects their different dataset composition.

\textbf{Paired comparisons yield 348 additional complete recoveries.}
QueryFlow completes 434 support sets missed by GFM-RAG, with 86 transitions in the reverse direction, for a net gain of 348 questions (Figure~\ref{fig:evidence-recovery}(b)). FR@10 weighted by question count rises from 62.90\% to 70.47\%, a gain of 7.57 percentage points. Net gains occur on every benchmark, comprising 121 questions on 2WikiMultiHopQA, 75 on HotpotQA, 56 on MuSiQue, and 96 on ConcurrentQA. Each additional recovery makes an entire support set available within the same document budget. The paired outcomes thus identify additional questions with complete support across all four collections.

\textbf{Additional recoveries preserve most previously complete support sets.}
Of the 1,705 questions incomplete under GFM-RAG, QueryFlow completes 434, or 25.5\%. It also retains complete recovery for 2,805 of the 2,891 questions completed by GFM-RAG, or 97.0\%. These rates pool questions across datasets; the main tables weight datasets equally. These conditional rates separate the recovery of missing support from the preservation of existing coverage. Together, they show that the net gain combines additional complete sets with substantial overlap in successful questions, rather than merely exchanging which questions are recovered.

\subsection{Retrieval Budget Analysis}
\label{sec:retrieval-budget-analysis}

\begin{figure}[t]
    \centering
    \includegraphics[width=\columnwidth,trim=0 1bp 0 3bp,clip]{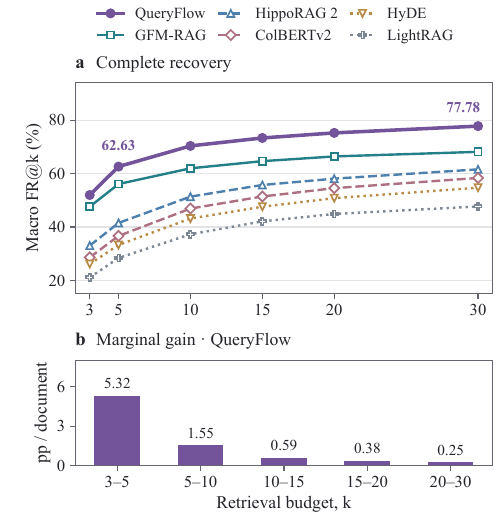}
    \caption{Retrieval budget analysis. (a) Macro FR@$k$ for methods that rank documents. (b) QueryFlow's marginal FR gain, $\Delta\mathrm{FR}/\Delta k$, in percentage points per added document.}
    \Description{Panel a shows six full recall curves at cutoffs 3, 5, 10, 15, 20, and 30. QueryFlow rises from 51.99 to 77.78 percent. Panel b shows marginal gains of 5.32, 1.55, 0.59, 0.38, and 0.25 percentage points per added document over successive budget intervals.}
    \label{fig:retrieval-budget}
\end{figure}

Figure~\ref{fig:retrieval-budget} compares complete evidence recovery as the retrieval budget increases from 3 to 30 documents across the four benchmarks. The curves address the following question.
\par\nopagebreak[4]

\noindent\textbf{RQ6.} \emph{How does retrieval budget affect complete recovery of evidence sets across methods?}

\textbf{QueryFlow reaches comparable recovery levels with fewer returned documents.}
QueryFlow reaches 62.63\% FR@5, compared with GFM-RAG's 61.98\% FR@10; its 70.38\% FR@10 also exceeds GFM-RAG's 68.15\% FR@30. At matched cutoffs, the margin over GFM-RAG grows from 4.37 points at $k=3$ to 8.40 at $k=10$ and 9.63 at $k=30$. QueryFlow also scores above HippoRAG~2, ColBERTv2, HyDE, and LightRAG throughout the tested range. These comparisons show that complete support sets appear earlier in QueryFlow's ranking. Aggregating relevance over connected entities can promote complementary documents together, making complete evidence available under smaller document budgets.

\textbf{Larger budgets recover additional complete support sets.}
QueryFlow's macro FR@$k$ increases from 51.99\% at $k=3$ to 70.38\% at $k=10$ and 77.78\% at $k=30$. Expanding the budget from 10 to 20 adds 4.87 percentage points, followed by another 2.54 points from 20 to 30. The gap between R@$k$ and FR@$k$ also narrows from 13.85 points at $k=5$ to 9.34 at $k=30$. Under a fixed ranking, a question becomes fully recovered when its last required document enters the returned prefix. Average gains per added document decline over the tested intervals in Figure~\ref{fig:retrieval-budget}(b), characterizing the tradeoff between retrieval budget and evidence completeness. The cutoff controls evidence passed to the generator without changing the ranking or repeating graph propagation. The paired and budget analyses address complementary aspects of evidence recovery. Paired outcomes show which questions gain or lose complete support at a fixed cutoff, whereas budget curves show how early complete support appears in a ranking. Together, they connect the aggregate retrieval improvements to the availability of a complete supporting set for individual questions.

\subsection{Efficiency Analysis}
\label{sec:efficiency-analysis}

Table~\ref{tab:efficiency-results} compares QueryFlow with existing methods in indexing cost, online time, and token consumption on 2WikiMultiHopQA. Together with answer quality in Table~\ref{tab:answer-results}, these results address the following question.

\par\nopagebreak[4]

\noindent\textbf{RQ7.} \emph{How does QueryFlow compare with existing methods in efficiency, token cost, and answer quality?}

\begin{table}[tbp]
    \centering
    \caption{Efficiency on 2WikiMultiHopQA. Online time and tokens include query processing, retrieval, and answer generation.}
    \label{tab:efficiency-results}
    \small
    \renewcommand{\arraystretch}{1.45}
    \setlength{\tabcolsep}{2pt}
    \begin{tabular*}{\columnwidth}{@{\extracolsep{\fill}}lrrrr@{}}
        \toprule
        \multirow{2}{*}{Method} & \multicolumn{2}{c}{Index/Build} & \multicolumn{2}{c}{Retrieval/Reasoning} \\
        \cmidrule(lr){2-3}\cmidrule(lr){4-5}
        & Time (ks) & Tokens (M) & Time (ks) & Tokens (M) \\
        \midrule
        GFM-RAG & 1.80 & 11.2 & 5.96 & 5.44 \\
        HippoRAG 2 & 2.08 & 11.1 & 1.36 & 8.53 \\
        ToG & 2.11 & 5.15 & 1.19 & 12.7 \\
        FastToG & 0.128 & 6.44 & 0.00820 & 1.25 \\
        \midrule
        \textbf{QueryFlow} & 1.80 & 11.2 & 0.580 & 3.59 \\
        \bottomrule
    \end{tabular*}
\end{table}

\textbf{Efficient online retrieval and answer generation.}
QueryFlow records 0.580 ks for retrieval and answer generation, compared with 5.96 ks for GFM-RAG, 1.36 ks for HippoRAG~2, and 1.19 ks for ToG (Table~\ref{tab:efficiency-results}). The corresponding reductions in recorded time are 90.3\%, 57.3\%, and 51.2\%, respectively. Query evolution refines the retrieval condition through tensor operations on the existing graph, keeping successive feedback steps within graph computation.

\textbf{Query adaptation adds no API-token cost.}
QueryFlow uses 3.59M tokens for retrieval and reasoning, compared with GFM-RAG's 5.44M, a reduction of 34.1\%. With comparable indexing costs of about 11.2M tokens, QueryFlow's estimated total is 14.8M versus 16.6M for GFM-RAG, a reduction of 10.8\%. The graph retriever itself consumes no API tokens, so updating the latent query state adds no intermediate query-generation token cost. This allows adaptive retrieval without repeated textual reasoning calls during graph propagation.

\textbf{Answer quality improves alongside resource efficiency.}
Table~\ref{tab:answer-results} reports LLM-Acc of 0.829 for QueryFlow and 0.668 for GFM-RAG on 2WikiMultiHopQA, a gain of 16.1 percentage points. The macro gain over GFM-RAG is 11.6 points. These answer results accompany lower recorded time and token costs relative to GFM-RAG in Table~\ref{tab:efficiency-results}. QueryFlow also achieves higher LLM-Acc than FastToG (0.829 versus 0.326), with FastToG offering a lower recorded online time and token cost.

\section{Conclusion}

We presented QueryFlow, a graph retrieval framework that supports multi-hop question answering through coupled evidence propagation and query evolution. Accumulated graph context updates a shared query state, which controls subsequent boundary inputs and relation features. Across four multi-hop QA benchmarks, QueryFlow improves macro R@10, FR@10, and LLM-Acc over GFM-RAG by 4.98, 8.40, and 11.6 percentage points, respectively. At a budget of five documents, it also exceeds this baseline's full recall at ten documents. Macro LLM-Acc also improves by 8.0 points over LinearRAG. Query-state interventions and interface ablations support the contribution of feedback within QueryFlow. The results connect evolving query conditions to more complete evidence sets and stronger downstream answers.

\clearpage
\bibliographystyle{ACM-Reference-Format}
\bibliography{references}

@inproceedings{cho-etal-2014-learning,
    title = "Learning Phrase Representations using {RNN} Encoder{--}Decoder for Statistical Machine Translation",
    author = {Cho, Kyunghyun  and
      van Merri{\"e}nboer, Bart  and
      Gulcehre, Caglar  and
      Bahdanau, Dzmitry  and
      Bougares, Fethi  and
      Schwenk, Holger  and
      Bengio, Yoshua},

    booktitle = "Proceedings of the 2014 Conference on Empirical Methods in Natural Language Processing ({EMNLP})",
    year = "2014",
    publisher = "Association for Computational Linguistics",
    pages = "1724--1734"
}

@inproceedings{trivedi-etal-2023-interleaving,
    title = "Interleaving Retrieval with Chain-of-Thought Reasoning for Knowledge-Intensive Multi-Step Questions",
    author = "Trivedi, Harsh  and
      Balasubramanian, Niranjan  and
      Khot, Tushar  and
      Sabharwal, Ashish",

    booktitle = "Proceedings of the 61st Annual Meeting of the Association for Computational Linguistics (Volume 1: Long Papers)",
    year = "2023",
    publisher = "Association for Computational Linguistics",
    pages = "10014--10037"
}

@inproceedings{yang-etal-2018-hotpotqa,
    title = "{H}otpot{QA}: A Dataset for Diverse, Explainable Multi-hop Question Answering",
    author = "Yang, Zhilin  and
      Qi, Peng  and
      Zhang, Saizheng  and
      Bengio, Yoshua  and
      Cohen, William  and
      Salakhutdinov, Ruslan  and
      Manning, Christopher D.",

    booktitle = "Proceedings of the 2018 Conference on Empirical Methods in Natural Language Processing",
    year = "2018",
    publisher = "Association for Computational Linguistics",
    pages = "2369--2380"
}

@inproceedings{ho-etal-2020-constructing,
    title = "Constructing A Multi-hop {QA} Dataset for Comprehensive Evaluation of Reasoning Steps",
    author = "Ho, Xanh  and
      Duong Nguyen, Anh-Khoa  and
      Sugawara, Saku  and
      Aizawa, Akiko",

    booktitle = "Proceedings of the 28th International Conference on Computational Linguistics",
    year = "2020",
    publisher = "International Committee on Computational Linguistics",
    pages = "6609--6625"
}

@article{trivedi-etal-2022-musique,
    title = "{M}u{S}i{Q}ue: Multihop Questions via Single-hop Question Composition",
    author = "Trivedi, Harsh  and
      Balasubramanian, Niranjan  and
      Khot, Tushar  and
      Sabharwal, Ashish",

    journal = "Transactions of the Association for Computational Linguistics",
    volume = "10",
    year = "2022",
    publisher = "MIT Press",
    pages = "539--554"
}

@inproceedings{zhu2021nbfnet,
 author = {Zhu, Zhaocheng and Zhang, Zuobai and Xhonneux, Louis-Pascal and Tang, Jian},
 booktitle = {Advances in Neural Information Processing Systems},

 pages = {29476--29490},
 publisher = {Curran Associates, Inc.},
 title = {Neural Bellman-Ford Networks: A General Graph Neural Network Framework for Link Prediction},
 volume = {34},
 year = {2021}
}

@inproceedings{xiong2021mdr,
  author = {Wenhan Xiong and Xiang Lorraine Li and Srinivasan Iyer and Jingfei Du and Patrick Lewis and William Wang and Yashar Mehdad and Yih, Wen-tau and Sebastian Riedel and Douwe Kiela and Barlas O{\u{g}}uz},
  title = {Answering Complex Open-Domain Questions with Multi-Hop Dense Retrieval},
  booktitle = {International Conference on Learning Representations},
  year = {2021},
}

@inproceedings{luo2025gfmrag,
  author = {Linhao Luo and Zicheng Zhao and Gholamreza Haffari and Chen Gong and Dinh Phung and Shirui Pan},
  title = {{GFM-RAG}: Graph Foundation Model for Retrieval Augmented Generation},
  booktitle = {Advances in Neural Information Processing Systems},
  volume = {38},
  pages = {36371--36405},
  publisher = {Curran Associates, Inc.},
  year = {2025},
}

@inproceedings{yang2015distmult,
  author = {Bishan Yang and Yih, Wen-tau and Xiaodong He and Jianfeng Gao and Li Deng},
  title = {Embedding Entities and Relations for Learning and Inference in Knowledge Bases},
  booktitle = {International Conference on Learning Representations},
  year = {2015},
}

@inproceedings{carbonell1998mmr,
  author = {Jaime Carbonell and Jade Goldstein},
  title = {The Use of {MMR}, Diversity-Based Reranking for Reordering Documents and Producing Summaries},
  booktitle = {Proceedings of the 21st Annual International ACM SIGIR Conference on Research and Development in Information Retrieval},
  year = {1998},
  publisher = {Association for Computing Machinery},
  pages = {335--336},
}

@article{concurrentqa,
  author       = {Simran Arora and
                  Patrick Lewis and
                  Angela Fan and
                  Jacob Kahn and
                  Christopher R{\'{e}}},
  title        = {Reasoning over Public and Private Data in Retrieval-Based Systems},
  journal      = {Transactions of the Association for Computational Linguistics},
  volume       = {11},
  pages        = {902--921},
  year         = {2023},
}

@inproceedings{raptor,
  author       = {Parth Sarthi and
                  Salman Abdullah and
                  Aditi Tuli and
                  Shubh Khanna and
                  Anna Goldie and
                  Christopher D. Manning},
  title        = {{RAPTOR:} Recursive Abstractive Processing for Tree-Organized Retrieval},
  booktitle    = {The Twelfth International Conference on Learning Representations},
  publisher    = {OpenReview.net},
  year         = {2024},
}

@inproceedings{hyde,
  author       = {Luyu Gao and
                  Xueguang Ma and
                  Jimmy Lin and
                  Jamie Callan},
  title        = {Precise Zero-Shot Dense Retrieval without Relevance Labels},
  booktitle    = {Proceedings of the 61st Annual Meeting of the Association for
                  Computational Linguistics (Volume 1: Long Papers)},
  pages        = {1762--1777},
  publisher    = {Association for Computational Linguistics},
  year         = {2023},
}

@inproceedings{colbertv2,
  author       = {Keshav Santhanam and
                  Omar Khattab and
                  Jon Saad{-}Falcon and
                  Christopher Potts and
                  Matei Zaharia},
  title        = {{ColBERTv2}: Effective and Efficient Retrieval via Lightweight Late
                  Interaction},
  booktitle    = {Proceedings of the 2022 Conference of the North American Chapter of
                  the Association for Computational Linguistics: Human Language
                  Technologies},
  pages        = {3715--3734},
  publisher    = {Association for Computational Linguistics},
  year         = {2022},
}

@inproceedings{g-retriever,
  author       = {Xiaoxin He and
                  Yijun Tian and
                  Yifei Sun and
                  Nitesh V. Chawla and
                  Thomas Laurent and
                  Yann LeCun and
                  Xavier Bresson and
                  Bryan Hooi},

  title        = {G-Retriever: Retrieval-Augmented Generation for Textual Graph Understanding
                  and Question Answering},
  booktitle    = {Advances in Neural Information Processing Systems},
  volume       = {37},
  pages        = {132876--132907},
  publisher    = {Curran Associates, Inc.},
  year         = {2024},
}

@inproceedings{lightrag,
  author       = {Zirui Guo and
                  Lianghao Xia and
                  Yanhua Yu and
                  Tu Ao and
                  Chao Huang},
  title        = {{\mbox{LightRAG}}: Simple and Fast Retrieval-Augmented Generation},
  booktitle    = {Findings of the Association for Computational Linguistics: {EMNLP} 2025},
  pages        = {10746--10761},
  publisher    = {Association for Computational Linguistics},
  year         = {2025},
}

@misc{graphrag,
  author       = {Darren Edge and
                  Ha Trinh and
                  Newman Cheng and
                  Joshua Bradley and
                  Alex Chao and
                  Apurva Mody and
                  Steven Truitt and
                  Dasha Metropolitansky and
                  Robert Osazuwa Ness and
                  Jonathan Larson},
  title        = {From Local to Global: {A} Graph {RAG} Approach to Query-Focused Summarization},
  year         = {2024},
  eprinttype   = {arXiv},
  eprint       = {2404.16130}
}

@inproceedings{dalk,
  author       = {Dawei Li and
                  Shu Yang and
                  Zhen Tan and
                  Jae Young Baik and
                  Sukwon Yun and
                  Joseph Lee and
                  Aaron Chacko and
                  Bojian Hou and
                  Duy Duong{-}Tran and
                  Ying Ding and
                  Huan Liu and
                  Li Shen and
                  Tianlong Chen},
  title        = {{DALK:} Dynamic Co-Augmentation of LLMs and {KG} to answer
                  Alzheimer's Disease Questions with Scientific Literature},
  booktitle    = {Findings of the Association for Computational Linguistics: {EMNLP} 2024},
  pages        = {2187--2205},
  publisher    = {Association for Computational Linguistics},
  year         = {2024},
}

@inproceedings{hipporag2,
  author       = {Bernal Jim{\'{e}}nez Guti{\'{e}}rrez and
                  Yiheng Shu and
                  Weijian Qi and
                  Sizhe Zhou and
                  Yu Su},
  title        = {From {RAG} to Memory: Non-Parametric Continual Learning for Large
                  Language Models},
  booktitle    = {Proceedings of the 42nd International Conference on Machine Learning},
  series       = {Proceedings of Machine Learning Research},
  volume       = {267},
  pages        = {21497--21515},
  publisher    = {PMLR},
  year         = {2025},
}

@inproceedings{linearrag,
  author = {Luyao Zhuang and Shengyuan Chen and Yilin Xiao and Huachi Zhou and Yujing Zhang and Hao Chen and Qinggang Zhang and Xiao Huang},
  title = {{LinearRAG}: Linear Graph Retrieval Augmented Generation on Large-scale Corpora},
  booktitle = {The Fourteenth International Conference on Learning Representations},
  year = {2026},
}

@article{fasttog,
  author       = {Xujian Liang and
                  Zhaoquan Gu},
  title        = {Fast Think-on-Graph: Wider, Deeper and Faster Reasoning of Large
                  Language Model on Knowledge Graph},
  journal      = {Proceedings of the AAAI Conference on Artificial Intelligence},
  volume       = {39},
  number       = {23},
  pages        = {24558--24566},
  year         = {2025},
}

@article{pathrag,
  author       = {Boyu Chen and
                  Zirui Guo and
                  Zidan Yang and
                  Yuluo Chen and
                  Junze Chen and
                  Zhenghao Liu and
                  Chuan Shi and
                  Cheng Yang},
  title        = {{PathRAG}: Pruning Graph-based Retrieval Augmented Generation with
                  Relational Paths},
  journal      = {Proceedings of the AAAI Conference on Artificial Intelligence},
  volume       = {40},
  number       = {36},
  pages        = {30183--30191},
  year         = {2026},
}

@inproceedings{jiang-etal-2023-active,
  title = {Active Retrieval Augmented Generation},
  author = {Jiang, Zhengbao and Xu, Frank and Gao, Luyu and Sun, Zhiqing and
            Liu, Qian and Dwivedi-Yu, Jane and Yang, Yiming and Callan, Jamie and Neubig, Graham},

  booktitle = {Proceedings of the 2023 Conference on Empirical Methods in Natural Language Processing},
  year = {2023},
  publisher = {Association for Computational Linguistics},
  pages = {7969--7992}
}

@inproceedings{asai2024selfrag,
  author = {Akari Asai and Zeqiu Wu and Yizhong Wang and Avirup Sil and Hannaneh Hajishirzi},
  title = {{Self-RAG}: Learning to Retrieve, Generate, and Critique through Self-Reflection},
  booktitle = {The Twelfth International Conference on Learning Representations},
  year = {2024},
}

@inproceedings{jeong-etal-2024-adaptive,
  title = {Adaptive-{RAG}: Learning to Adapt Retrieval-Augmented Large Language Models through Question Complexity},
  author = {Jeong, Soyeong and Baek, Jinheon and Cho, Sukmin and Hwang, Sung Ju and Park, Jong},

  booktitle = {Proceedings of the 2024 Conference of the North American Chapter of the Association for Computational Linguistics: Human Language Technologies (Volume 1: Long Papers)},
  year = {2024},
  publisher = {Association for Computational Linguistics},
  pages = {7036--7050}
}

@inproceedings{hipporag,
  author = {Guti\'{e}rrez, Bernal Jim\'{e}nez and Shu, Yiheng and Gu, Yu and Yasunaga, Michihiro and Su, Yu},
  title = {{HippoRAG}: Neurobiologically Inspired Long-Term Memory for Large Language Models},
  booktitle = {Advances in Neural Information Processing Systems},

  volume = {37},
  pages = {59532--59569},
  publisher = {Curran Associates, Inc.},
  year = {2024},
}

@inproceedings{zhu-etal-2025-knowledge,
  title = {Knowledge Graph-Guided Retrieval Augmented Generation},
  author = {Zhu, Xiangrong and Xie, Yuexiang and Liu, Yi and Li, Yaliang and Hu, Wei},

  booktitle = {Proceedings of the 2025 Conference of the Nations of the Americas Chapter of the Association for Computational Linguistics: Human Language Technologies (Volume 1: Long Papers)},
  year = {2025},
  publisher = {Association for Computational Linguistics},
  pages = {8912--8924},
}

@inproceedings{sun2024tog,
  author = {Jiashuo Sun and Chengjin Xu and Lumingyuan Tang and Saizhuo Wang and Chen Lin and
            Yeyun Gong and Lionel M. Ni and Heung-Yeung Shum and Jian Guo},
  title = {{Think-on-Graph}: Deep and Responsible Reasoning of Large Language Model on Knowledge Graph},
  booktitle = {The Twelfth International Conference on Learning Representations},
  year = {2024},
}

@inproceedings{luo2024rog,
  author = {Linhao Luo and Yuan-Fang Li and Gholamreza Haffari and Shirui Pan},
  title = {Reasoning on Graphs: Faithful and Interpretable Large Language Model Reasoning},
  booktitle = {The Twelfth International Conference on Learning Representations},
  year = {2024},
}

@inproceedings{galkin2024ultra,
  author = {Mikhail Galkin and Xinyu Yuan and Hesham Mostafa and Jian Tang and Zhaocheng Zhu},
  title = {Towards Foundation Models for Knowledge Graph Reasoning},
  booktitle = {The Twelfth International Conference on Learning Representations},
  year = {2024},
}

@inproceedings{lewis2020rag,
  author = {Patrick Lewis and Ethan Perez and Aleksandra Piktus and Fabio Petroni and
            Vladimir Karpukhin and Naman Goyal and Heinrich K{\"u}ttler and Mike Lewis and
            Yih, Wen-tau and Tim Rockt{\"a}schel and Sebastian Riedel and Douwe Kiela},
  title = {Retrieval-Augmented Generation for Knowledge-Intensive {NLP} Tasks},
  booktitle = {Advances in Neural Information Processing Systems},
  volume = {33},
  pages = {9459--9474},
  publisher = {Curran Associates, Inc.},
  year = {2020},
}

@inproceedings{karpukhin2020dpr,
  author = {Vladimir Karpukhin and Barlas Oguz and Sewon Min and Patrick Lewis and
            Ledell Wu and Sergey Edunov and Danqi Chen and Yih, Wen-tau},
  title = {Dense Passage Retrieval for Open-Domain Question Answering},
  booktitle = {Proceedings of the 2020 Conference on Empirical Methods in Natural Language Processing (EMNLP)},
  year = {2020},
  pages = {6769--6781},
  publisher = {Association for Computational Linguistics},
}

@inproceedings{guu2020realm,
  author = {Kelvin Guu and Kenton Lee and Zora Tung and Panupong Pasupat and Ming-Wei Chang},
  title = {{REALM}: Retrieval-Augmented Language Model Pre-Training},
  booktitle = {Proceedings of the 37th International Conference on Machine Learning},
  series = {Proceedings of Machine Learning Research},
  volume = {119},
  year = {2020},
  pages = {3929--3938},
  publisher = {PMLR},
}

@inproceedings{izacard2021fid,
  author = {Gautier Izacard and Edouard Grave},
  title = {Leveraging Passage Retrieval with Generative Models for Open Domain Question Answering},
  booktitle = {Proceedings of the 16th Conference of the European Chapter of the Association for Computational Linguistics: Main Volume},
  year = {2021},
  pages = {874--880},
  publisher = {Association for Computational Linguistics},
}

@article{izacard2023atlas,
  author = {Gautier Izacard and Patrick Lewis and Maria Lomeli and Lucas Hosseini and
            Fabio Petroni and Timo Schick and Jane Dwivedi-Yu and Armand Joulin and
            Sebastian Riedel and Edouard Grave},
  title = {Atlas: Few-shot Learning with Retrieval Augmented Language Models},
  journal = {Journal of Machine Learning Research},
  volume = {24},
  number = {251},
  year = {2023},
  pages = {1--43},
}

@inproceedings{borgeaud2022retro,
  author = {Sebastian Borgeaud and Arthur Mensch and Jordan Hoffmann and Trevor Cai and
            Eliza Rutherford and Katie Millican and George van den Driessche and Jean-Baptiste Lespiau and
            Bogdan Damoc and Aidan Clark and Diego de Las Casas and Aurelia Guy and Jacob Menick and
            Roman Ring and Tom Hennigan and Saffron Huang and Loren Maggiore and Chris Jones and
            Albin Cassirer and Andy Brock and Michela Paganini and Geoffrey Irving and Oriol Vinyals and
            Simon Osindero and Karen Simonyan and Jack W. Rae and Erich Elsen and Laurent Sifre},
  title = {Improving Language Models by Retrieving from Trillions of Tokens},
  booktitle = {Proceedings of the 39th International Conference on Machine Learning},
  series = {Proceedings of Machine Learning Research},
  volume = {162},
  year = {2022},
  pages = {2206--2240},
  publisher = {PMLR},
}

@article{izacard2022contriever,
  author = {Gautier Izacard and Mathilde Caron and Lucas Hosseini and Sebastian Riedel and
            Piotr Bojanowski and Armand Joulin and Edouard Grave},
  title = {Unsupervised Dense Information Retrieval with Contrastive Learning},
  journal = {Transactions on Machine Learning Research},
  year = {2022},
}

@inproceedings{khattab2020colbert,
  author = {Omar Khattab and Matei Zaharia},
  title = {{ColBERT}: Efficient and Effective Passage Search via Contextualized Late Interaction over {BERT}},
  booktitle = {Proceedings of the 43rd International ACM SIGIR Conference on Research and Development in Information Retrieval},
  year = {2020},
  pages = {39--48},
  publisher = {Association for Computing Machinery},
}

@inproceedings{qi-etal-2019-answering,
    title = "Answering Complex Open-domain Questions Through Iterative Query Generation",
    author = "Qi, Peng  and
      Lin, Xiaowen  and
      Mehr, Leo  and
      Wang, Zijian  and
      Manning, Christopher D.",

    booktitle = "Proceedings of the 2019 Conference on Empirical Methods in Natural Language Processing and the 9th International Joint Conference on Natural Language Processing ({EMNLP}-{IJCNLP})",
    year = "2019",
    publisher = "Association for Computational Linguistics",
    pages = "2590--2602"
}

@inproceedings{khattab2021baleen,
 author = {Khattab, Omar and Potts, Christopher and Zaharia, Matei},
 booktitle = {Advances in Neural Information Processing Systems},

 pages = {27670--27682},
 publisher = {Curran Associates, Inc.},
 title = {Baleen: Robust Multi-Hop Reasoning at Scale via Condensed Retrieval},
 volume = {34},
 year = {2021}
}

@inproceedings{zhang-etal-2024-end,
    title = "End-to-End Beam Retrieval for Multi-Hop Question Answering",
    author = "Zhang, Jiahao  and
      Zhang, Haiyang  and
      Zhang, Dongmei  and
      Liu, Yong  and
      Huang, Shen",

    booktitle = "Proceedings of the 2024 Conference of the {N}orth {A}merican Chapter of the {A}ssociation for {C}omputational {L}inguistics: Human Language Technologies (Volume 1: Long Papers)",
    year = "2024",
    publisher = "Association for Computational Linguistics",
    pages = "1718--1731"
}

@inproceedings{ma-etal-2023-query,
    title = "Query Rewriting in Retrieval-Augmented Large Language Models",
    author = "Ma, Xinbei  and
      Gong, Yeyun  and
      He, Pengcheng  and
      Zhao, Hai  and
      Duan, Nan",

    booktitle = "Proceedings of the 2023 Conference on Empirical Methods in Natural Language Processing",
    year = "2023",
    publisher = "Association for Computational Linguistics",
    pages = "5303--5315"
}

@inproceedings{shi-etal-2024-replug,
    title = "{REPLUG}: Retrieval-Augmented Black-Box Language Models",
    author = "Shi, Weijia  and
      Min, Sewon  and
      Yasunaga, Michihiro  and
      Seo, Minjoon  and
      James, Richard  and
      Lewis, Mike  and
      Zettlemoyer, Luke  and
      Yih, Wen-tau",

    booktitle = "Proceedings of the 2024 Conference of the {N}orth {A}merican Chapter of the {A}ssociation for {C}omputational {L}inguistics: Human Language Technologies (Volume 1: Long Papers)",
    year = "2024",
    publisher = "Association for Computational Linguistics",
    pages = "8371--8384"
}

@inproceedings{shao-etal-2023-enhancing,
    title = "Enhancing Retrieval-Augmented Large Language Models with Iterative Retrieval-Generation Synergy",
    author = "Shao, Zhihong  and
      Gong, Yeyun  and
      Shen, Yelong  and
      Huang, Minlie  and
      Duan, Nan  and
      Chen, Weizhu",

    booktitle = "Findings of the {A}ssociation for {C}omputational {L}inguistics: {EMNLP} 2023",
    year = "2023",
    publisher = "Association for Computational Linguistics",
    pages = "9248--9274"
}

@inproceedings{su-etal-2024-dragin,
    title = "{DRAGIN}: Dynamic Retrieval Augmented Generation based on the Real-time Information Needs of Large Language Models",
    author = "Su, Weihang  and
      Tang, Yichen  and
      Ai, Qingyao  and
      Wu, Zhijing  and
      Liu, Yiqun",

    booktitle = "Proceedings of the 62nd Annual Meeting of the {A}ssociation for {C}omputational {L}inguistics (Volume 1: Long Papers)",
    year = "2024",
    publisher = "Association for Computational Linguistics",
    pages = "12991--13013"
}

@inproceedings{hu2025grag,
  author = {Yuntong Hu and Zhihan Lei and Zheng Zhang and Bo Pan and Chen Ling and Liang Zhao},
  title = {{GRAG}: Graph Retrieval-Augmented Generation},
  booktitle = {Findings of the Association for Computational Linguistics: NAACL 2025},
  year = {2025},
  pages = {4145--4157},
  publisher = {Association for Computational Linguistics},
}

@inproceedings{mavromatis2025gnnrag,
  author = {Costas Mavromatis and George Karypis},
  title = {{GNN-RAG}: Graph Neural Retrieval for Efficient Large Language Model Reasoning on Knowledge Graphs},
  booktitle = {Findings of the Association for Computational Linguistics: ACL 2025},
  year = {2025},
  pages = {16682--16699},
  publisher = {Association for Computational Linguistics},
}

@inproceedings{li2025subgraphrag,
  author = {Mufei Li and Siqi Miao and Pan Li},
  title = {Simple is Effective: The Roles of Graphs and Large Language Models in Knowledge-Graph-Based Retrieval-Augmented Generation},
  booktitle = {The Thirteenth International Conference on Learning Representations},
  year = {2025},
}

@inproceedings{luo2026graphr1,
  author = {Haoran Luo and Haihong E and Guanting Chen and Qika Lin and Yikai Guo and Fangzhi Xu and Zemin Kuang and Meina Song and Xiaobao Wu and Yifan Zhu and Anh Tuan Luu},
  title = {{Graph-R1}: Towards Agentic {GraphRAG} Framework via End-to-end Reinforcement Learning},
  booktitle = {Proceedings of the 43rd International Conference on Machine Learning},
  series = {Proceedings of Machine Learning Research},
  volume = {306},
  pages = {82220--82240},
  year = {2026},
  publisher = {PMLR},
}

@inproceedings{sun2019pullnet,
  author = {Haitian Sun and Tania Bedrax-Weiss and William W. Cohen},
  title = {{PullNet}: Open Domain Question Answering with Iterative Retrieval on Knowledge Bases and Text},
  booktitle = {Proceedings of the 2019 Conference on Empirical Methods in Natural Language Processing and the 9th International Joint Conference on Natural Language Processing (EMNLP-IJCNLP)},
  year = {2019},
  pages = {2380--2390},
  publisher = {Association for Computational Linguistics},
}

@inproceedings{zhu2023astarnet,
  author = {Zhaocheng Zhu and Xinyu Yuan and Mikhail Galkin and Sophie Xhonneux and Ming Zhang and Maxime Gazeau and Jian Tang},
  title = {{A*Net}: A Scalable Path-based Reasoning Approach for Knowledge Graphs},
  booktitle = {Advances in Neural Information Processing Systems},
  volume = {36},
  pages = {59323--59336},
  year = {2023},
  publisher = {Curran Associates, Inc.},
}

@inproceedings{li2025gnnret,
  author = {Zijian Li and Qingyan Guo and Jiawei Shao and Lei Song and Jiang Bian and Jun Zhang and Rui Wang},
  title = {Graph Neural Network Enhanced Retrieval for Question Answering of Large Language Models},
  booktitle = {Proceedings of the 2025 Conference of the Nations of the Americas Chapter of the Association for Computational Linguistics: Human Language Technologies (Volume 1: Long Papers)},
  year = {2025},
  pages = {6612--6633},
  publisher = {Association for Computational Linguistics},
}

@inproceedings{talmor-berant-2018-web,
  author = {Alon Talmor and Jonathan Berant},
  title = {The Web as a Knowledge-Base for Answering Complex Questions},
  booktitle = {Proceedings of the 2018 Conference of the North American Chapter of the Association for Computational Linguistics: Human Language Technologies, Volume 1 (Long Papers)},
  year = {2018},
  pages = {641--651},
  publisher = {Association for Computational Linguistics},
}

@inproceedings{jiang-etal-2020-hover,
  author = {Yichen Jiang and Shikha Bordia and Zheng Zhong and Charles Dognin and Maneesh Singh and Mohit Bansal},
  title = {{HoVer}: A Dataset for Many-Hop Fact Extraction And Claim Verification},
  booktitle = {Findings of the Association for Computational Linguistics: EMNLP 2020},
  year = {2020},
  pages = {3441--3460},
  publisher = {Association for Computational Linguistics},
}

@inproceedings{gao-etal-2025-rag,
  author = {Guangze Gao and Zixuan Li and Chunfeng Yuan and Jiawei Li and Jianzhuo Wu and Yuehao Zhang and Xiaolong Jin and Bing Li and Weiming Hu},
  title = {{D-RAG}: Differentiable Retrieval-Augmented Generation for Knowledge Graph Question Answering},
  booktitle = {Proceedings of the 2025 Conference on Empirical Methods in Natural Language Processing},
  year = {2025},
  pages = {35398--35417},
  publisher = {Association for Computational Linguistics},
}

@article{huang2026relink,
  author = {Manzong Huang and Chenyang Bu and Yi He and Xingrui Zhuo and Xindong Wu},
  title = {{Relink}: Constructing Query-Driven Evidence Graph On-the-Fly for {GraphRAG}},
  journal = {Proceedings of the AAAI Conference on Artificial Intelligence},
  volume = {40},
  number = {37},
  pages = {31202--31210},
  year = {2026},
}

@inproceedings{zhuo2025effective,
  author = {Xingrui Zhuo and Jiapu Wang and Gongqing Wu and Shirui Pan and Xindong Wu},
  title = {Effective Instruction Parsing Plugin for Complex Logical Query Answering on Knowledge Graphs},
  booktitle = {Proceedings of the ACM on Web Conference 2025},
  pages = {4780--4792},
  year = {2025},
  publisher = {ACM},
  doi = {10.1145/3696410.3714794},
}

@inproceedings{ijcai2025p413,
  author = {Xingrui Zhuo and Shirui Pan and Jiapu Wang and Gongqing Wu and Zan Zhang and Rui Li and Zizhong Wei and Xindong Wu},
  title = {Progressive Prefix-Memory Tuning for Complex Logical Query Answering on Knowledge Graphs},
  booktitle = {Proceedings of the Thirty-Fourth International Joint Conference on Artificial Intelligence},
  pages = {3716--3724},
  year = {2025},
  publisher = {International Joint Conferences on Artificial Intelligence Organization},
  doi = {10.24963/ijcai.2025/413},
  url = {https://www.ijcai.org/proceedings/2025/413},
}

@inproceedings{wang2024llmda,
  author = {Jiapu Wang and Kai Sun and Linhao Luo and Wei Wei and Yongli Hu and Alan Liew and Shirui Pan and Baocai Yin},
  title = {Large Language Models-guided Dynamic Adaptation for Temporal Knowledge Graph Reasoning},
  booktitle = {Advances in Neural Information Processing Systems},
  volume = {37},
  pages = {8384--8410},
  year = {2024},
  publisher = {Neural Information Processing Systems Foundation},
  doi = {10.52202/079017-0269},
}

@inproceedings{zhuo2026krlm,
  author = {Xingrui Zhuo and Jiapu Wang and Gongqing Wu and Zhongyuan Wang and Jichen Zhang and Shirui Pan and Xindong Wu},
  title = {Knowledge Reasoning Language Model: Unifying Knowledge and Language for Inductive Knowledge Graph Reasoning},
  booktitle = {International Conference on Learning Representations},
  year = {2026},
  url = {https://proceedings.iclr.cc/paper_files/paper/2026/file/c6e9f9d49cc052879efd19da0c3e1900-Paper-Conference.pdf},
}

@article{wang2023mcrl,
  author = {Jiapu Wang and Boyue Wang and Junbin Gao and Yongli Hu and Baocai Yin},
  title = {Multi-Concept Representation Learning for Knowledge Graph Completion},
  journal = {ACM Transactions on Knowledge Discovery from Data},
  volume = {17},
  number = {1},
  pages = {1--19},
  year = {2023},
  publisher = {Association for Computing Machinery},
  doi = {10.1145/3533017},
}

@inproceedings{wang2024ime,
  author = {Jiapu Wang and Zheng Cui and Boyue Wang and Shirui Pan and Junbin Gao and Baocai Yin and Wen Gao},
  title = {{IME}: Integrating Multi-curvature Shared and Specific Embedding for Temporal Knowledge Graph Completion},
  booktitle = {Proceedings of the ACM Web Conference 2024},
  pages = {1954--1962},
  year = {2024},
  publisher = {Association for Computing Machinery},
  doi = {10.1145/3589334.3645361},
}

@article{wang2023tdn,
  author = {Jiapu Wang and Boyue Wang and Junbin Gao and Xiaoyan Li and Yongli Hu and Baocai Yin},
  title = {{TDN}: Triplet Distributor Network for Knowledge Graph Completion},
  journal = {IEEE Transactions on Knowledge and Data Engineering},
  volume = {35},
  number = {12},
  pages = {13002--13014},
  year = {2023},
  publisher = {Institute of Electrical and Electronics Engineers},
  doi = {10.1109/TKDE.2023.3272568},
}

@article{wang2024mli,
  author = {Jiapu Wang and Boyue Wang and Junbin Gao and Simin Hu and Yongli Hu and Baocai Yin},
  title = {Multi-Level Interaction Based Knowledge Graph Completion},
  journal = {IEEE/ACM Transactions on Audio, Speech, and Language Processing},
  volume = {32},
  pages = {386--396},
  year = {2024},
  publisher = {Institute of Electrical and Electronics Engineers},
  doi = {10.1109/TASLP.2023.3331121},
}

@article{wang2024qdn,
  author = {Jiapu Wang and Boyue Wang and Junbin Gao and Xiaoyan Li and Yongli Hu and Baocai Yin},
  title = {{QDN}: A Quadruplet Distributor Network for Temporal Knowledge Graph Completion},
  journal = {IEEE Transactions on Neural Networks and Learning Systems},
  volume = {35},
  number = {10},
  pages = {14018--14030},
  year = {2024},
  publisher = {Institute of Electrical and Electronics Engineers},
  doi = {10.1109/TNNLS.2023.3274230},
}

@article{wang2026comgrl,
  author = {Jinlu Wang and Yanfeng Sun and Jiapu Wang and Junbin Gao and Shaofan Wang and Baocai Yin and Jipeng Guo},
  title = {{ComGRL}: Comprehensive Graph Representation Learning from Local to Global Bridged by Mixup},
  journal = {IEEE Transactions on Computational Social Systems},
  volume = {13},
  number = {3},
  pages = {2839--2852},
  year = {2026},
  publisher = {Institute of Electrical and Electronics Engineers},
  doi = {10.1109/TCSS.2026.3665450},
}

@inproceedings{zhao2025ragc,
  author = {Tianxiang Zhao and Youqing Wang and Jinlu Wang and Jiapu Wang and Mingliang Cui and Junbin Gao and Jipeng Guo},
  title = {Hybrid-Collaborative Augmentation and Contrastive Sample Adaptive-Differential Awareness for Robust Attributed Graph Clustering},
  booktitle = {Advances in Neural Information Processing Systems},
  volume = {38},
  pages = {128567--128592},
  year = {2025},
  publisher = {Neural Information Processing Systems Foundation},
  doi = {10.52202/085713-3867},
}
\clearpage

\appendix
\section{Additional Experimental Details}
\label{app:additional-results}

\subsection{Experimental Setup}
\label{app:experimental-setup}

\paragraph{2WikiMultiHopQA}
This benchmark combines Wikidata relations with Wikipedia text to construct questions with explicit multi-hop reasoning paths~\cite{ho-etal-2020-constructing}. These annotations specify the intermediate relations linking a question to its answer, providing a structured basis for assessing evidence-set recovery.

\paragraph{HotpotQA}
Questions draw on multiple Wikipedia documents and include comparisons between entities~\cite{yang-etal-2018-hotpotqa}. Sentence-level supporting facts identify the evidence needed for an answer. We map these annotations to source documents to define the target evidence sets.

\paragraph{MuSiQue}
MuSiQue composes single-hop questions into connected two- to four-hop questions~\cite{trivedi-etal-2022-musique}. Intermediate answers link successive steps, reducing opportunities for disconnected reasoning and making subsequent evidence needs depend on earlier discoveries. We use the answerable examples.

\paragraph{ConcurrentQA}
ConcurrentQA requires evidence from distinct public and private data distributions~\cite{concurrentqa}. We use the supplied documents as a shared retrieval collection to evaluate evidence recovery across heterogeneous sources with the same document-ranking interface.

\paragraph{Evaluation subsets.}
Following GFM-RAG~\cite{luo2025gfmrag}, LinearRAG~\cite{linearrag}, and Relink~\cite{huang2026relink}, we use fixed evaluation subsets. Matching LinearRAG's reference question lists to source records yields 1,000 questions each for 2WikiMultiHopQA and HotpotQA, with HotpotQA drawn from distractor validation. For MuSiQue, we match the reference list to records in the answerable validation split. HotpotQA and MuSiQue are shuffled with seed 42 before matching; list membership determines selection. ConcurrentQA uses all 1,600 validation questions. Input sets are shared across methods. Retrieval metrics are averaged over questions with nonempty support sets (Section~\ref{sec:exp-settings}).

\paragraph{Retrieval corpora.}
Supporting and distractor documents from the selected examples form each benchmark's retrieval corpus. Table~\ref{tab:dataset-statistics} reports selected input records and corpus sizes after preprocessing. Each question is evaluated against this collection, with $k$ limiting the number of returned documents. Original support annotations define the target evidence sets.

\begin{table}[tbp]
    \centering
    \caption{Evaluation input records and retrieval corpus sizes.}
    \label{tab:dataset-statistics}
    \small
    \begin{tabular*}{\columnwidth}{@{\extracolsep{\fill}}lrr@{}}
        \toprule
        Dataset & Input records & Documents \\
        \midrule
        2WikiMultiHopQA & 1,000 & 9,594 \\
        HotpotQA & 1,000 & 9,811 \\
        MuSiQue & 1,003 & 11,666 \\
        ConcurrentQA & 1,600 & 2,840 \\
        \bottomrule
    \end{tabular*}
\end{table}

\paragraph{Training data and supervision.}
We use 20,000 examples and their associated content from each of 2WikiMultiHopQA, HotpotQA, and MuSiQue, totaling 60,000 training examples. Entity labels $y_i$ are derived from these training datasets. The graph interface is defined in Section~\ref{sec:formulation}. Extraction, normalization, linking, grounding, and document-to-entity label construction are detailed in the supplementary code.

\subsection{Implementation Details}
\label{app:implementation-settings}

\paragraph{Model architecture.}
The frozen \texttt{all-mpnet-base-v2} encoder supplies precomputed 768-dimensional question and relation embeddings, linearly projected to $d=512$. The propagation operator uses the NBFNet formulation~\cite{zhu2021nbfnet}. The retriever uses $L=6$ layers with 512 hidden units, DistMult messages, sum aggregation, residual connections, layer normalization, and ReLU. The query controller also has 512-dimensional states. The full model enables both feedback interfaces with $\alpha=1$, attention temperature 1, and query-update dropout 0.

\paragraph{Optimization and hardware.}
We train the retriever and query controller from random initialization on a single NVIDIA A100 using AdamW, learning rate $10^{-5}$, batch size one, gradient norm clipping at 1.0, and seed 1024. Activation checkpointing reduces propagation memory; the text encoder remains frozen.

\paragraph{Answer generation and judgment.}
All methods generate answers with Qwen3.5-9B. Following LinearRAG~\cite{linearrag}, LLM-Acc records LLM judge acceptance; Contain checks for the normalized gold answer. Prompts, context handling, decoding, judge configuration, and output parsing are detailed in the supplementary code.

\subsection{Additional Results}

Tables~\ref{tab:queryflow-answer-results} and~\ref{tab:retrieval-results-appendix} report additional answer metrics and retrieval scores by dataset. Macro averages weight datasets equally.

\begin{table}[htbp]
    \centering
    \caption{Additional answer generation metrics for QueryFlow (\%). Macro is the unweighted mean across datasets.}
    \label{tab:queryflow-answer-results}
    \small
    \setlength{\tabcolsep}{2pt}
    \resizebox{\columnwidth}{!}{%
    \begin{tabular}{@{}lrrrrrr@{}}
        \toprule
        Dataset & EM & F1 & Precision & Recall & Contain & LLM-Acc \\
        \midrule
        2WikiMultiHopQA & 65.3\% & 75.4\% & 73.9\% & 83.3\% & 81.9\% & 82.9\% \\
        HotpotQA & 56.7\% & 69.6\% & 71.0\% & 77.0\% & 69.5\% & 78.6\% \\
        MuSiQue & 24.5\% & 35.1\% & 35.3\% & 49.0\% & 37.5\% & 43.9\% \\
        ConcurrentQA & 55.7\% & 63.5\% & 63.5\% & 72.5\% & 67.8\% & 73.2\% \\
        \midrule
        Macro & 50.5\% & 60.9\% & 60.9\% & 70.4\% & 64.2\% & 69.6\% \\
        \bottomrule
    \end{tabular}
    }%
\end{table}

\begin{table*}[t]
    \centering
    \caption{Retrieval scores for each dataset (\%), ordered by macro FR@30. We compare R@$k$ and FR@$k$ for methods with document-level outputs, obtained directly or by mapping retrieved graph or hierarchical units to source documents. Macro is the unweighted mean across datasets. Best in \textbf{bold}, second \underline{underlined}.}
    \label{tab:retrieval-results-appendix}
    \footnotesize
    \renewcommand{\arraystretch}{1.25}
    \setlength{\tabcolsep}{1.8pt}
    \resizebox{\textwidth}{!}{%
    \begin{tabular}{l*{5}{rrrr}}
        \toprule
        & \multicolumn{4}{c}{2WikiMultiHopQA}
        & \multicolumn{4}{c}{HotpotQA}
        & \multicolumn{4}{c}{MuSiQue}
        & \multicolumn{4}{c}{ConcurrentQA}
        & \multicolumn{4}{c}{Macro} \\
        \cmidrule(lr){2-5}\cmidrule(lr){6-9}\cmidrule(lr){10-13}
        \cmidrule(lr){14-17}\cmidrule(lr){18-21}
        Method & R@10 & FR@10 & R@30 & FR@30 & R@10 & FR@10 & R@30 & FR@30
        & R@10 & FR@10 & R@30 & FR@30 & R@10 & FR@10 & R@30 & FR@30
        & R@10 & FR@10 & R@30 & FR@30 \\
        \midrule
        G-Retriever & 32.10 & 11.28 & 34.37 & 12.08 & 44.85 & 23.51 & 46.57 & 24.12 & 25.80 & 11.60 & 27.72 & 12.10 & 42.14 & 20.50 & 45.66 & 22.70 & 36.22 & 16.72 & 38.58 & 17.75 \\
        DALK & 63.68 & 28.30 & 63.95 & 28.80 & 63.78 & 40.54 & 67.40 & 45.57 & 40.44 & 10.87 & 44.35 & 14.08 & 60.47 & 37.79 & 66.81 & 43.91 & 57.09 & 29.38 & 60.63 & 33.09 \\
        RAPTOR & 59.22 & 27.10 & 80.30 & 61.70 & 65.80 & 40.70 & 65.80 & 40.70 & 40.87 & 8.87 & 66.43 & 38.29 & 60.22 & 34.69 & 60.22 & 34.69 & 56.53 & 27.84 & 68.19 & 43.84 \\
        LightRAG & 52.20 & 24.50 & 59.75 & 31.80 & 82.35 & 67.60 & 88.90 & 79.30 & 57.40 & 24.23 & 69.42 & 39.38 & 54.69 & 32.81 & 60.59 & 40.31 & 61.66 & 37.28 & 69.67 & 47.70 \\
        GraphRAG & 70.62 & 39.90 & 74.35 & 46.50 & 77.60 & 60.80 & 86.70 & 76.90 & 49.20 & 14.56 & 60.10 & 26.12 & 65.44 & 51.06 & 77.44 & 67.56 & 65.72 & 41.58 & 74.65 & 54.27 \\
        HyDE & 58.52 & 30.80 & 66.22 & 38.80 & 86.05 & 73.80 & 91.75 & 84.60 & \underline{63.17} & 31.01 & \textbf{74.53} & \underline{46.46} & 58.34 & 37.06 & 68.16 & 49.06 & 66.52 & 43.17 & 75.16 & 54.73 \\
        ColBERTv2 & 70.50 & 42.20 & 74.98 & 49.10 & 86.75 & 74.30 & 92.70 & 85.50 & 56.60 & 22.33 & 67.95 & 37.19 & 70.97 & 48.81 & 78.91 & 61.69 & 71.20 & 46.91 & 78.63 & 58.37 \\
        HippoRAG 2 & 73.82 & 45.59 & 77.66 & 51.70 & \underline{89.32} & \underline{79.84} & \underline{93.38} & \underline{87.36} & \textbf{64.42} & 31.11 & \underline{74.51} & 45.56 & 69.60 & 48.97 & 78.27 & 61.74 & 74.29 & 51.38 & 80.95 & 61.59 \\
        GFM-RAG & \underline{88.52} & \underline{75.00} & \underline{89.72} & \underline{77.50} & 86.19 & 76.68 & 89.69 & 82.38 & 57.04 & \underline{31.84} & 65.32 & 42.32 & \underline{76.06} & \underline{64.39} & \underline{80.91} & \underline{70.40} & \underline{76.95} & \underline{61.98} & \underline{81.41} & \underline{68.15} \\
        \midrule
        QueryFlow & \textbf{93.81} & \textbf{87.60} & \textbf{95.36} & \textbf{90.70} & \textbf{90.99} & \textbf{84.78} & \textbf{94.74} & \textbf{90.99} & 61.60 & \textbf{37.82} & 72.27 & \textbf{51.20} & \textbf{81.33} & \textbf{71.29} & \textbf{86.09} & \textbf{78.24} & \textbf{81.93} & \textbf{70.38} & \textbf{87.12} & \textbf{77.78} \\
        \bottomrule
    \end{tabular}
    }
\end{table*}

\section{Training Objective and Retrieval Algorithm}
\label{app:implementation}

\subsection{Document Evidence Readout}
\label{app:readout}

Default entity scoring uses $q_0$ during both training and retrieval. Intermediate query updates influence document ranking through the final entity representations $H_L$.

\paragraph{Aggregation over all entities.}
Softplus preserves entity ordering and yields positive contributions $\mathbf v$. With $f_i=\sum_j A_{ij}$ and $w_i^{\mathrm{doc}}=1/f_i$ for $f_i>0$ (zero otherwise), the document scores are
\begin{equation}
    \mathbf d=A^\top\!\left(\mathbf w^{\mathrm{doc}}\odot\mathbf v\right).
    \label{eq:matrix-readout}
\end{equation}
Here $A\in\{0,1\}^{|V|\times|D|}$ and $\mathbf d\in\mathbb R^{|D|}$. All entities contribute; the top $k$ scores define $\widehat D_k(x)$. Paired propagation interventions keep the scorer and readout fixed.

\subsection{Entity-Level Training Objective}
\label{app:training}

Training masks $y\in\{0,1\}^{|V|}$ define positives $P=\{i:y_i=1\}$ and negatives $N=\{i:y_i=0\}$. Binary and listwise losses supervise the graph and query controller.

\paragraph{Entity relevance objective.}
Let $p_i=\sigma(s_i)$, where $s_i$ is computed using the anchored scorer in Eq.~\eqref{eq:entity-score}. The binary cross-entropy for entity $i$ is
\begin{equation}
    \ell_i=-y_i\log p_i-(1-y_i)\log(1-p_i).
    \label{eq:bce-entity}
\end{equation}
Positive weights are uniform; negative weights depend on current logits,
\begin{equation}
    \omega_i=
    \begin{cases}
        |P|^{-1}, & i\in P,\\
        \displaystyle\operatorname{sg}\!
            \left(\frac{\exp(s_i/T)}
                        {\sum_{j\in N}\exp(s_j/T)}\right), & i\in N.
    \end{cases}
    \label{eq:adversarial-weights}
\end{equation}
Here $T=0.2$, and $\operatorname{sg}$ stops gradients through the negative weights. Negative entities with higher scores receive greater weight, concentrating the classification loss on negatives that the retriever currently ranks highly. The weighted binary loss is
\begin{equation}
    \mathcal L_{\mathrm{BCE}}
       =\frac{\sum_{i\in V}\omega_i\ell_i}
                    {\sum_{i\in V}\omega_i}.
    \label{eq:bce}
\end{equation}

ListCE (Eq.~\eqref{eq:listce}, $\varepsilon=10^{-5}$) normalizes sigmoid probabilities over entities and averages negative log mass over all positives, excluding empty targets. The combined loss in Eq.~\eqref{eq:loss} is averaged over questions in each batch.

Aggregation of entity scores into document scores and document ranking are performed at inference time, outside the entity-level training objective. The query controller is trained through its influence on subsequent graph propagation and final entity scores; the supporting-document sets define the downstream evaluation targets.

\subsection{Retrieval Procedure}
\label{app:algorithm}

Algorithm~\ref{alg:queryflow} applies $L$ propagation layers with query updates between successive layers before ranking documents. The default $\alpha=1$ applies the gated increment before normalization; $\alpha=0$ fixes the query without changing learned propagation parameters.

At layer $l$, propagation first computes $H_l$ under $q_{l-1}$. When $l<L$, the updated entity states are summarized to form $q_l$, which conditions the next layer. The default scorer combines $H_L$ with $q_0$. The $q_L$ variant adds the terminal update defined in Section~\ref{sec:transition} and substitutes $q_L$ in the scorer. Graph structure, grounding, and document associations remain fixed.

\begin{algorithm}[htbp]
    \caption{QueryFlow with the default $q_0$ readout}
    \label{alg:queryflow}
    \begin{algorithmic}[1]
        \Require $G,A,m_x,e_x,\{e_r\}$; parameters; $L,\alpha,k$
        \Ensure Ranked top-$k$ documents $\widehat D_k(x)$
        \State $q_0\gets\phi_q(e_x)$; $R^0(r)\gets\phi_r(e_r)$ for all $r$
        \State $H_0(i)\gets m_x(i)q_0$ for all $i\in V$
        \For{$l=1,\ldots,L$}
            \State $B_l(i)\gets m_x(i)q_{l-1}$
            \State $\gamma_{l-1}\gets\sigma(W_rq_{l-1}+b_r)$
            \State $\bar R_l(r)\gets\psi_l(R^0(r)\odot\gamma_{l-1})$
            \State Compute $H_l$ using Eqs.~\eqref{eq:messages}--\eqref{eq:node-update}
            \If{$l<L$}
                \State Compute $z_l$ using Eqs.~\eqref{eq:attention}--\eqref{eq:summary}
                \If{$\alpha=0$}
                    \State $q_l\gets q_{l-1}$
                \Else
                    \State Form $\widetilde q_l,g_l,\Delta_l$ using Eq.~\eqref{eq:state-gate}
                    \State $q_l\gets\operatorname{LN}_q(q_{l-1}+\alpha\Delta_l)$
                \EndIf
            \EndIf
        \EndFor
        \State $s_i\gets\operatorname{MLP}([H_L(i);q_0])$ for all $i\in V$
        \State $v_i\gets\operatorname{softplus}(s_i)$ for all $i\in V$
        \State $f_i\gets\sum_{j=1}^{|D|} A_{ij}$ for all $i\in V$
        \State $w_i^{\mathrm{doc}}\gets 1/f_i$ if $f_i>0$, otherwise $0$
        \State $d_j\gets\sum_{i\in V}A_{ij}w_i^{\mathrm{doc}}v_i$ for all documents $j$
        \State \Return the $k$ highest-scoring documents
    \end{algorithmic}
\end{algorithm}

\section{Readout Analysis}
\label{app:readout-controls}

We compare aggregation over all entities with two truncated entity readouts, document-aware entity MMR~\cite{carbonell1998mmr} and relevance-based Top-20 selection. All three use the same propagated entity representations and reciprocal document frequency weighting.

\begin{table}[htbp]
    \centering
    \caption{Document readout comparison. Macro retrieval scores across four datasets (\%).}
    \label{tab:readout-controls}
    \small
    \setlength{\tabcolsep}{4pt}
    \begin{tabular}{@{}lrrrr@{}}
        \toprule
        Readout & R@10 & FR@10 & R@30 & FR@30 \\
        \midrule
        MMR Top-20 & 80.48 & 67.84 & 85.31 & 74.72 \\
        Top-20 (no MMR) & 80.57 & 67.79 & 84.87 & 73.90 \\
        \midrule
        All entities (default) & \textbf{81.93} & \textbf{70.38} & \textbf{87.12} & \textbf{77.78} \\
        \bottomrule
    \end{tabular}
\end{table}

\textbf{Aggregation over all entities improves complete evidence recovery.} The default readout exceeds both truncated alternatives on all four metrics in Table~\ref{tab:readout-controls}. Relative to MMR Top-20, the gains are 2.54 and 3.07 percentage points in FR@10 and FR@30. Aggregating every entity contribution lets documents receive relevance signals distributed beyond a small selected subset, providing a mechanism for retaining complementary evidence. Within the truncated alternatives, MMR improves R@30 and FR@30 by 0.44 and 0.82 points, with mixed but small differences at $k=10$ ($-0.09$ R@10 and $+0.05$ FR@10).

\textbf{Reciprocal weighting preserves each entity's total contribution.}
For $f_i>0$, summing $A_{ij}v_i/f_i$ over documents yields $v_i$. Summing the complete document score vector gives
\begin{equation}
    \sum_{j=1}^{|D|}d_j
    =\sum_{i\in V}w_i^{\mathrm{doc}}v_i\sum_{j=1}^{|D|}A_{ij}
    =\sum_{i\in V:f_i>0}v_i.
    \label{eq:readout-conservation}
\end{equation}
Document associations redistribute each entity's score across its source documents. An entity appearing in one document contributes its full score there; one appearing in several documents divides that contribution equally among them. This property controls how graph relevance transfers to documents without selecting a fixed number of entities. Each document can still accumulate contributions from multiple entities, so its rank depends on their combined relevance and document frequencies. The weighting acts on entity frequency rather than normalizing by document length.


\end{document}